\documentclass[10pt]{optica-article}

\journal{opticajournal}
\articletype{Research Article}

\usepackage{graphicx}
\usepackage{comment}
\usepackage{amsmath}
\usepackage{subcaption}
\usepackage{float}

\date{December 2025}

\begin{document}

\title{Simulating optical microcavities by inverting the thermal transfer function}

\author{Sudipta Nayak}

\address{Centre for Nano Science and Engineering, Indian Institute of Science, Bengaluru 560012, India}

\email{\authormark{*}nayaks@anl.gov}

\begin{abstract*}
I present a reduced-order approach for incorporating distributed thermal
dynamics into nonlinear optical-microcavity simulations without repeatedly
solving the time-dependent heat equation. A harmonic finite-element heat solve
is used to obtain the thermal transfer function, which is introduced into the
time domain through either a rational approximation or an impulse-response
convolution. The method is validated against heat-equation-coupled finite-element
simulations for prescribed TE$_1$-mode heating and for nonlinear silicon
microcavity dynamics. For the feed-forward TE$_1$ tests up to 100 MHz and simplified cavity dynamics, the error remains below 4\%. The reduced models reproduce
the local self-pulsation dynamics, although long-time phase alignment is
sensitive to the spatial definition of the thermal transfer function and to
impulse-response truncation. These results establish a physics-based route for
retaining multi-timescale thermal memory while substantially reducing the cost
of coupled optical--thermal simulations.
\end{abstract*}

\section{Introduction}

Thermal and free-carrier effects strongly influence the nonlinear dynamics of
silicon optical microcavities. These effects are commonly modeled using coupled
equations for the intracavity optical field, free-carrier density, and
temperature~\cite{pernice2010time,johnson2006self}. In many reduced-order cavity
models, the thermal response is approximated by a single first-order equation
with one effective thermal time constant~\cite{pernice2010time,johnson2006self,borghi2021modeling}.
Although useful, this approximation can fail when several thermal
modes~\cite{chandorkar2009multimode,pavlov2023microresonator} contribute
appreciably to the cavity response. This limitation is particularly relevant
for spatially extended optical structures, in which a high density of thermal
modes can contribute to the effective temperature~\cite{straube2011thermal,arendt2009weyl}.

Borghi \textit{et al.}~\cite{borghi2021modeling} demonstrated the limitations
of the conventional first-order model for certain cavity configurations and
proposed a phenomenological second-order equation that more accurately captures
the observed dynamics.

A more general description treats the cavity temperature as a distributed
thermal response driven by optical absorption. Here, finite-element simulations
of the harmonic heat equation are used to extract the cavity thermal transfer
function. This response is then incorporated into time-domain simulations using
either a rational approximation or an impulse-response convolution. Both
approaches retain the multi-timescale thermal response while avoiding solution
of the full heat equation during each nonlinear cavity simulation.

The method is validated against full finite-element time-domain simulations for
constant-power excitation, sinusoidal heating, and nonlinear self-pulsation in
a silicon microcavity. Both simplified and detailed carrier models are
considered. The reduced models reproduce the principal cavity dynamics while
reducing the repeated thermal calculation to either a finite set of internal
states or a discrete convolution. I identify two practical requirements for
accuracy: the reduced model must use the thermal transfer function associated
with the same spatially weighted temperature as the transient model, and the
impulse response must retain sufficient thermal memory when nonlinear feedback
is present.

\section{Thermal Transfer Function Model}

The temperature dynamics of the cavity are governed by the heat equation
\begin{equation}
\rho(\mathbf r)c_p(\mathbf r)
\frac{\partial \Delta T(\mathbf r,t)}{\partial t}
=
\nabla \cdot
\left[
\kappa(\mathbf r)\nabla \Delta T(\mathbf r,t)
\right]
+
Q(\mathbf r,t),
\label{eq:heat_equation_deltaT}
\end{equation}
where $\rho$, $c_p$, and $\kappa$ are the density, heat capacity, and thermal conductivity, respectively, and $Q(\mathbf r,t)$ is the absorbed-power density.

For a fixed optical mode profile, the heat source can be written as
\begin{equation}
Q(\mathbf r,t)=X(\mathbf r)P_{\mathrm{abs}}(t),
\label{eq:separable_source}
\end{equation}
where $X(\mathbf r)$ is the normalized spatial heating profile and
$P_{\mathrm{abs}}(t)$ is the absorbed optical power. For nonmagnetic dielectric
waveguides, I approximate the heating profile using the normalized optical
energy density,
\begin{equation}
X(\mathbf r)
=
\frac{
\epsilon(\mathbf r)|\mathbf E(\mathbf r)|^2
}{
\displaystyle
\int_{\Omega}
\epsilon(\mathbf r)|\mathbf E(\mathbf r)|^2
\,d^3\mathbf  r
}.
\label{eq:source_weight}
\end{equation}

Fourier transforming Eq.~\eqref{eq:heat_equation_deltaT} gives the harmonic thermal problem
\begin{equation}
\left[
i\omega C(\mathbf{r})
-
\nabla \cdot \kappa(\mathbf{r}) \nabla
\right]
H(\mathbf{r},\omega)
=
X(\mathbf{r}),
\label{eq:harmonic_ttf_pde}
\end{equation}
where $C(\mathbf r)=\rho(\mathbf r)c_p(\mathbf r)$ and
$H(\mathbf r,\omega)$ is the spatial thermal transfer function. The harmonic
temperature response is then
\begin{equation}
\widetilde{T}(\mathbf r,\omega)
=
H(\mathbf r,\omega)\widetilde{P}_{\mathrm{abs}}(\omega).
\label{eq:harmonic_temp_response}
\end{equation}

The scalar temperature relevant to the optical resonance shift is the optically
weighted effective temperature~\cite{johnson2006self},
\begin{equation}
T_{\mathrm{eff}}(t)
=
\frac{
\displaystyle \int_{\mathrm{Si}}
\Delta T(\mathbf r,t)\,
n_{\mathrm{Si}}^2
|\mathbf E(\mathbf r)|^2
\,d^3\mathbf r
}{
\displaystyle \int_{\mathrm{all}}
\varepsilon(\mathbf r)
|\mathbf E(\mathbf r)|^2
\,d^3\mathbf r
}.
\label{eq:johnson_teff}
\end{equation}
The corresponding effective thermal transfer function,
$H_{\mathrm{eff}}(\omega)$, is obtained by applying the same optical weighting
to $H(\mathbf r,\omega)$.
\begin{equation}
\widetilde{P}_{\mathrm{abs}}(\omega)H_{\mathrm{eff}}(\omega)
=
\frac{
\displaystyle \int_{\mathrm{Si}}
\widetilde{ T}(\mathbf r,\omega)\,
n_{\mathrm{Si}}^2
|\mathbf E(\mathbf r)|^2
\,d^3\mathbf r
}{
\displaystyle \int_{\mathrm{all}}
\varepsilon(\mathbf r)
|\mathbf E(\mathbf r)|^2
\,d^3\mathbf r
}.
\label{eq:johnson_Heff}
\end{equation}

Using a unit harmonic source sets
$\widetilde{P}_{\mathrm{abs}}(\omega)=1$. The resulting frequency-domain
transfer function provides the starting point for the reduced-order time-domain
models described below.

I also consider a point-sampled thermal response evaluated at the centroid of
the optical overlap integral, $\mathbf{r}_c$, defined by

\begin{equation}
\mathbf r_c
=
\frac{
\displaystyle \int_{\mathrm{Si}}
\mathbf r\,
n_{\mathrm{Si}}^2
|\mathbf E(\mathbf r)|^2
\,d^3\mathbf r
}{
\displaystyle \int_{\mathrm{Si}}
n_{\mathrm{Si}}^2
|\mathbf E(\mathbf r)|^2
\,d^3\mathbf r
}.
\label{eq:optical_centroid}
\end{equation}

\section{Reduced-Order Thermal Models}

\subsection{Rational-Approximation Method}

In the first approach, the effective thermal transfer function is approximated by a stable pole expansion,
\begin{equation}
H_{\mathrm{th}}(s)
\approx
\sum_{k=1}^{N_p}
\frac{c_k}{s+\lambda_k},
\qquad
\lambda_k>0 .
\label{eq:rational_fit}
\end{equation}
This representation converts the distributed thermal memory into a finite set of first-order state equations,
\begin{equation}
\frac{dz_k}{dt}
=
-\lambda_k z_k
+
c_k P_{\mathrm{abs}}(t),
\qquad
T_{\mathrm{eff}}(t)
=
\sum_{k=1}^{N_p}z_k(t).
\label{eq:pole_states}
\end{equation}
The rational approximation can therefore be coupled directly to the optical
and carrier-rate equations and integrated using standard stiff ODE solvers.

For a steady initial absorbed power $P_{\mathrm{abs},0}$, the thermal states are initialized as
\begin{equation}
z_k(0)=\frac{c_k}{\lambda_k}P_{\mathrm{abs},0}.
\label{eq:zk_init_power}
\end{equation}

\subsection{Impulse-Response Method}

In the second approach, the harmonic thermal transfer function is transformed
into a causal impulse response $h_{\mathrm{th}}(t)$. The effective temperature
is then computed by convolving the absorbed-power history with this response,
\begin{equation}
T_{\mathrm{eff}}(t)
=
\int_0^t
h_{\mathrm{th}}(t-t')
P_{\mathrm{abs}}(t')\,dt' .
\label{eq:impulse_conv}
\end{equation}
On a discrete simulation grid, this becomes
\begin{equation}
T_{\mathrm{eff}}[n]
=
\sum_{m=0}^{M}
K[m]P_{\mathrm{abs}}[n-m],
\label{eq:discrete_conv}
\end{equation}
where $K[m]$ is the finite convolution kernel obtained from the sampled impulse
response. This method avoids a rational-fitting step and provides a
direct time-domain representation of the finite-element thermal response. In
practice, the kernel must be truncated to a finite memory length; the effect of
this truncation is examined in Sec.~\ref{sec:impulse_convergence}. It should be noted that equation \ref {eq:impulse_conv} assumes that the initial temperature is 0 and there is no absorbed power history before $t=0$.

\section{Modal TE$_1$ Heating}
\label{sec:te1_heating}

\begin{figure}[H]
\centering
\captionsetup[subfigure]{skip=1pt}

\begin{subfigure}[t]{0.43\linewidth}
    \centering
    \includegraphics[width=0.95\linewidth]{figures/TE1}
    \caption{}
    \label{fig:te1_source}
\end{subfigure}
\begin{subfigure}[t]{0.43\linewidth}
    \centering
    \includegraphics[width=0.95\linewidth]{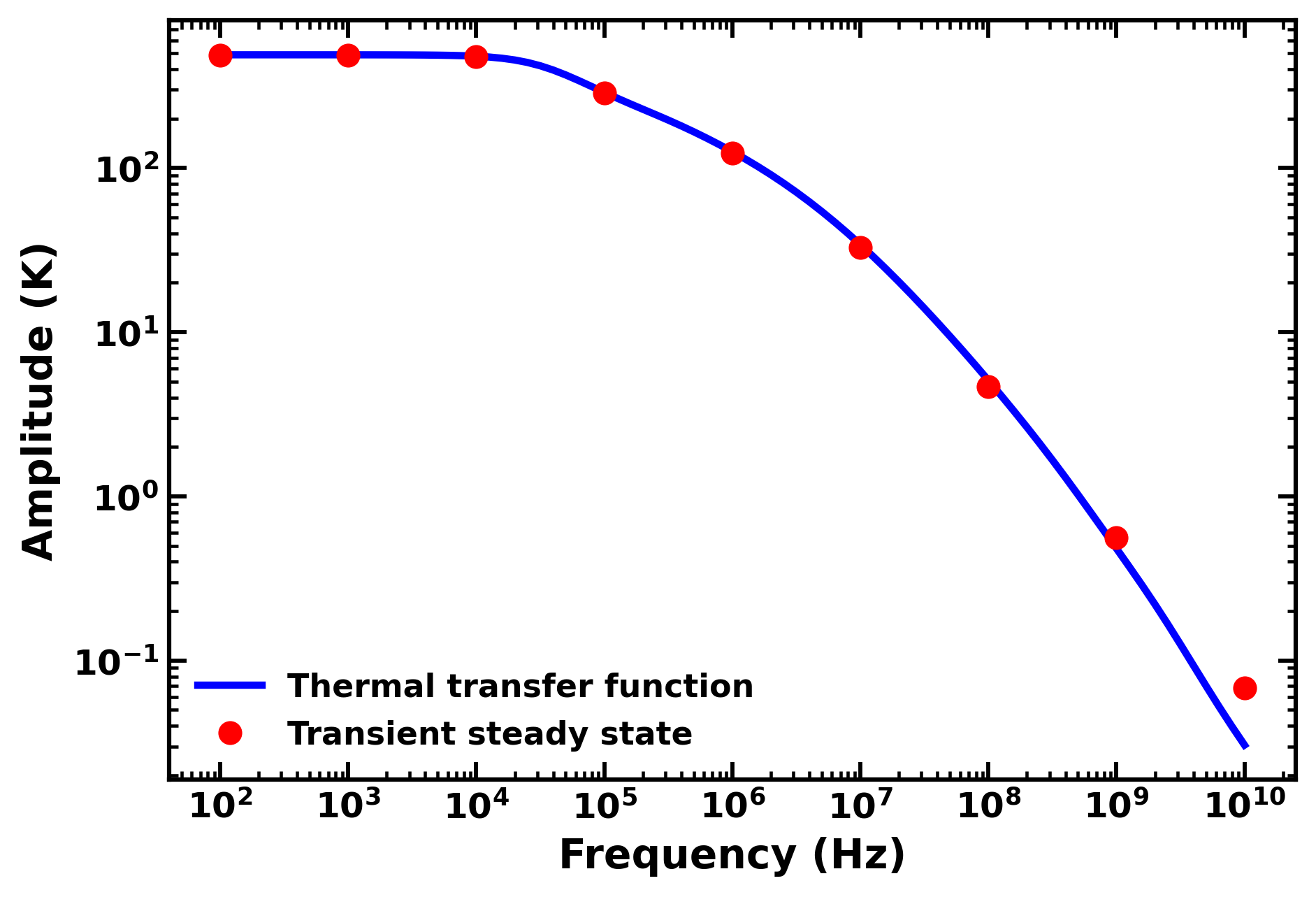}
    \caption{}
    \label{fig:te1_amp}
\end{subfigure}

\begin{subfigure}[t]{0.43\linewidth}
    \centering
    \includegraphics[width=0.95\linewidth]{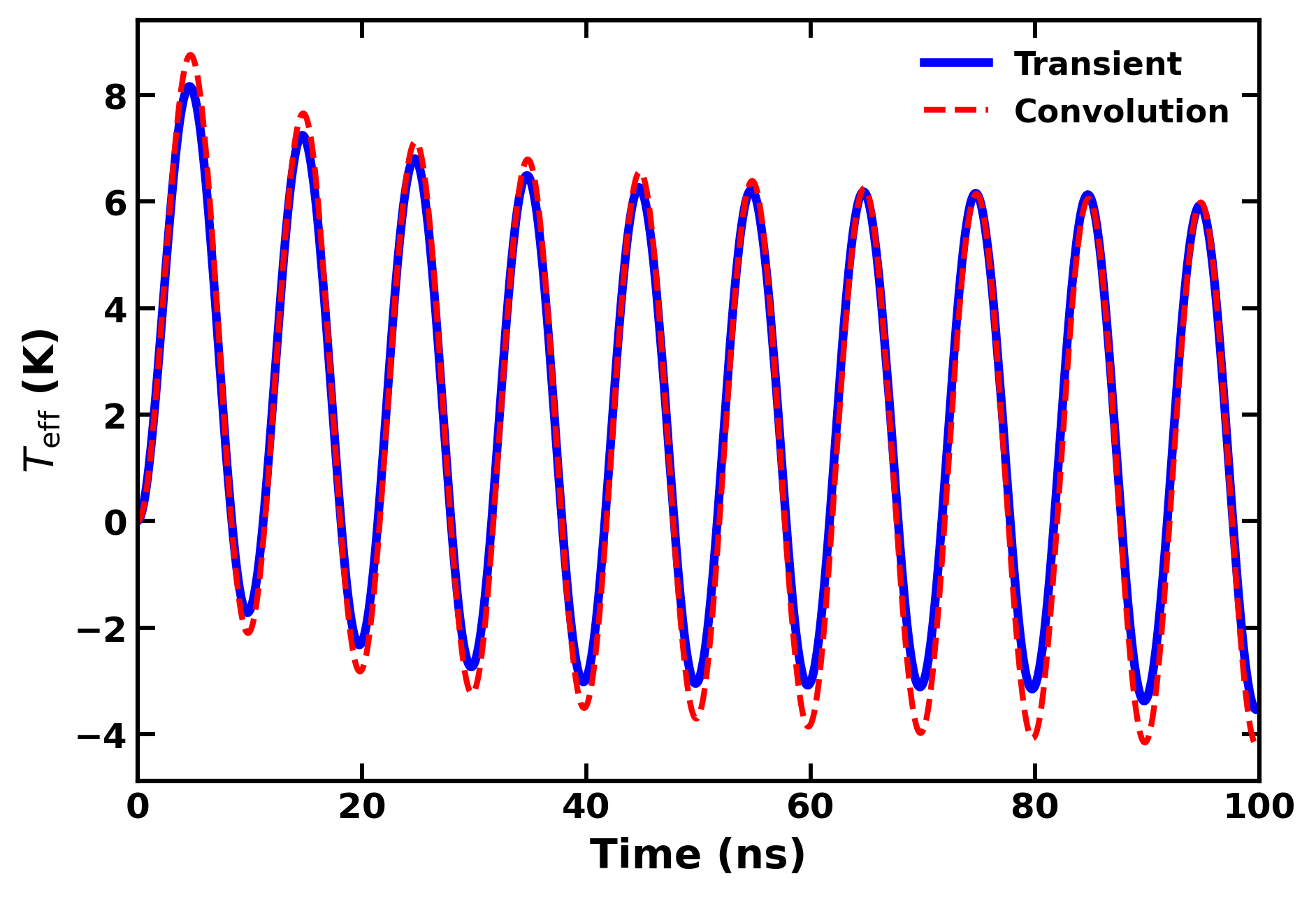}
    \caption{}
    \label{fig:te1_100mhz}
\end{subfigure}
\begin{subfigure}[t]{0.43\linewidth}
    \centering
    \includegraphics[width=0.95\linewidth]{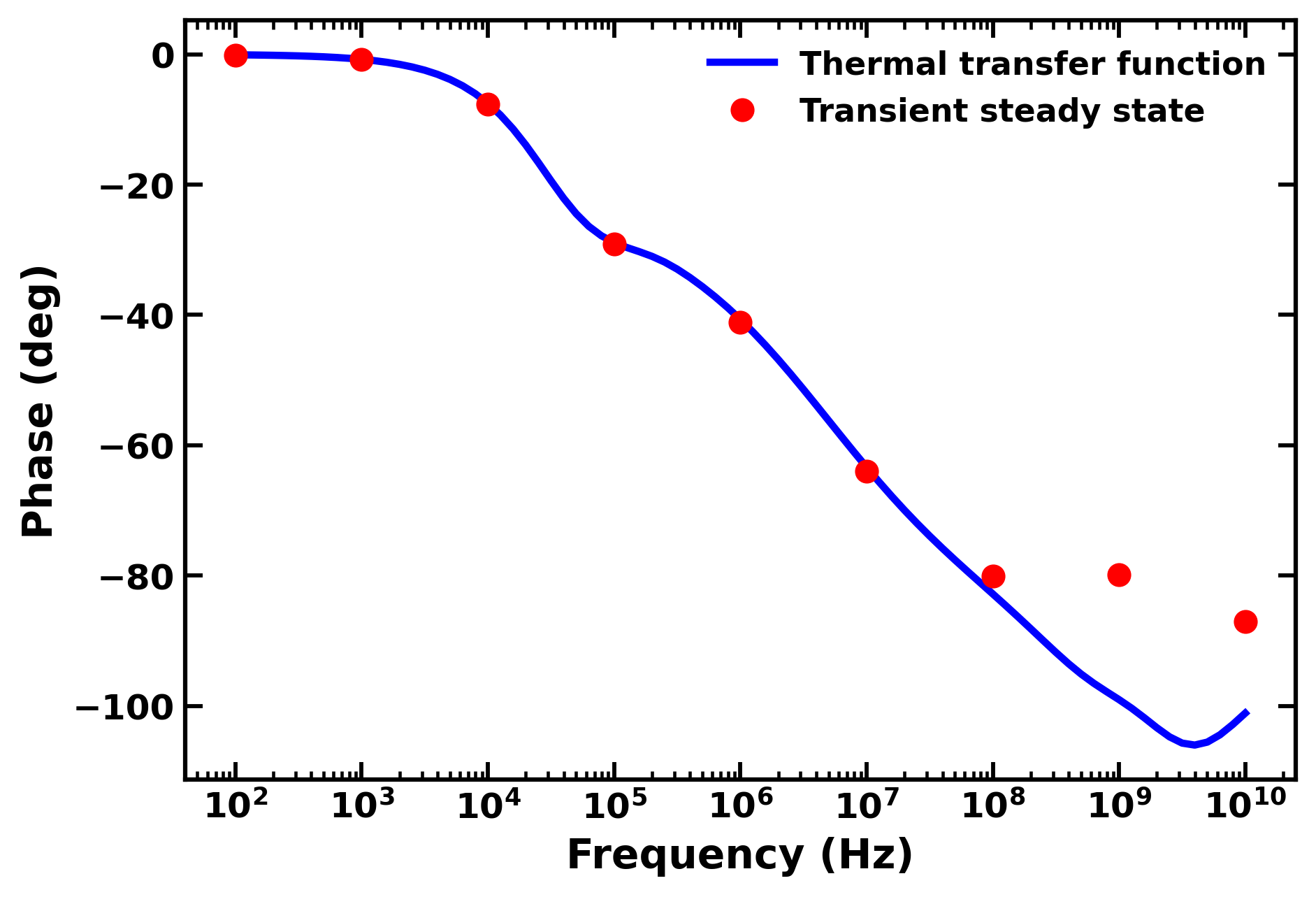}
    \caption{}
    \label{fig:te1_phase}
\end{subfigure}

\caption{
Validation of the impulse-response method for sinusoidal TE$_1$-mode heating.
The heat source is taken as
$Q(\mathbf r,t)=X_{\mathrm{TE1}}(\mathbf r)\sin(\omega t)$. The centroid-sampled
thermal transfer function is inverted to generate the convolution result,
whereas the COMSOL transient reference is the globally weighted effective
temperature.
(a) TE$_1$ heating profile and waveguide geometry.
(b) Amplitude of the centroid thermal transfer function.
(c) Time-domain comparison between the COMSOL transient solution and the
impulse-response convolution at $100~\mathrm{MHz}$.
(d) Phase of the centroid thermal transfer function.
}
\label{fig:te1_validation}
\end{figure}

\begin{table}[H]
\centering
\caption{Percentage temperature root-mean-square error (RMSE) between the
globally weighted transient simulation and the centroid-based impulse-response
convolution for
$Q(\mathbf r,t)=X_{\mathrm{TE1}}(\mathbf r)\sin(\omega t)$.}
\label{tab:te1_rmse}
\begin{tabular}{c c c c}
\hline
Frequency & $\mathrm{RMSE}_{T}$ (\%) & Frequency & $\mathrm{RMSE}_{T}$ (\%) \\
\hline
$100~\mathrm{Hz}$ & 3.223 & $1~\mathrm{MHz}$ & 0.223 \\
$1~\mathrm{kHz}$ & 3.223 & $10~\mathrm{MHz}$ & 1.248 \\
$10~\mathrm{kHz}$ & 3.093 & $100~\mathrm{MHz}$ & 3.377 \\
$100~\mathrm{kHz}$ & 1.288 & $1~\mathrm{GHz}$ & 12.108 \\
\hline
\end{tabular}
\end{table}

For the SOI rib waveguide (air/Si/SiO\textsubscript{2}) with a
\(1~\mu\text{m} \times 0.4~\mu\text{m}\) core, a \(150~\text{nm}\)-high
Si base on SiO\textsubscript{2}, and a ring radius of \(50~\mu\text{m}\),
the TE\(_1\) mode was used to define the optical absorption profile. The
harmonic heat equation was solved using the energy-density distribution
obtained from the TE\(_1\) mode. The resulting thermal-response amplitude
and phase sampled at the waveguide centroid are shown in
Figs.~\ref{fig:te1_amp} and \ref{fig:te1_phase}, respectively. To test whether
this centroid transfer function can approximate the globally weighted thermal
response, I invert it to obtain an impulse response and compare the resulting
temperature with a full transient heat simulation. The native time
discretization produced by the inverse fast Fourier transform is used, and the
impulse response is truncated at $15~\mu\mathrm{s}$ based on convergence of its
tail. The transient heat equation solved for this comparison is
\begin{equation}
\rho c_p \frac{\partial T(\mathbf{r},t)}{\partial t}
=
\nabla \cdot \left[\kappa \nabla T(\mathbf{r},t)\right]
+
X_{\mathrm{TE1}}(\mathbf{r})\sin(\omega t),
\label{eq:transient_heat}
\end{equation}
with the same TE\(_1\)-derived energy-density profile. The source represents a
signed harmonic heating perturbation about equilibrium. The time-domain
comparison at $100~\mathrm{MHz}$ is shown in
Fig.~\ref{fig:te1_100mhz}. The transient reference is the globally evaluated
effective temperature, whereas the convolution uses the transfer function
evaluated at the centroid.

The two methods agree closely up to $100~\mathrm{MHz}$. Beyond this,
differences emerge in both amplitude and phase, as shown in
Table~\ref{tab:te1_rmse}. For this table and subsequent comparisons, the
temperature RMSE is defined as
\[
\mathrm{RMSE}_{T,\%}
=
100
\frac{
\sqrt{\frac{1}{N}\sum_{i=1}^{N}
\left(T_{\mathrm{TTF},i}-T_{\mathrm{COMSOL},i}\right)^2}
}{
\sqrt{\frac{1}{N}\sum_{i=1}^{N}
T_{\mathrm{COMSOL},i}^{2}}
}.
\]
The larger error at $1~\mathrm{GHz}$ indicates the limit of the present
centroid-based impulse model. The most likely sources of deviation are the
difference between the point-sampled centroid response and the optically
weighted global temperature, together with the finite $15~\mu\mathrm{s}$
impulse-response window.

This feed-forward test shows that a centroid-sampled thermal transfer function
can approximate the globally weighted response over a broad frequency range.
However, the approximation becomes less accurate as the spatial thermal field
develops stronger high-frequency structure. It therefore does not imply that
the centroid transfer function can replace the optically weighted effective
transfer function in nonlinear cavity simulations. That distinction is examined
in Sec.~\ref{sec:sensitivity_ttf}.

\section{Cavity Simulations}
\label{sec:detailed_cavity_simulation}

\begin{figure}[htbp]
    \centering

    \begin{subfigure}{0.48\textwidth}
        \centering
        \includegraphics[width=\linewidth]{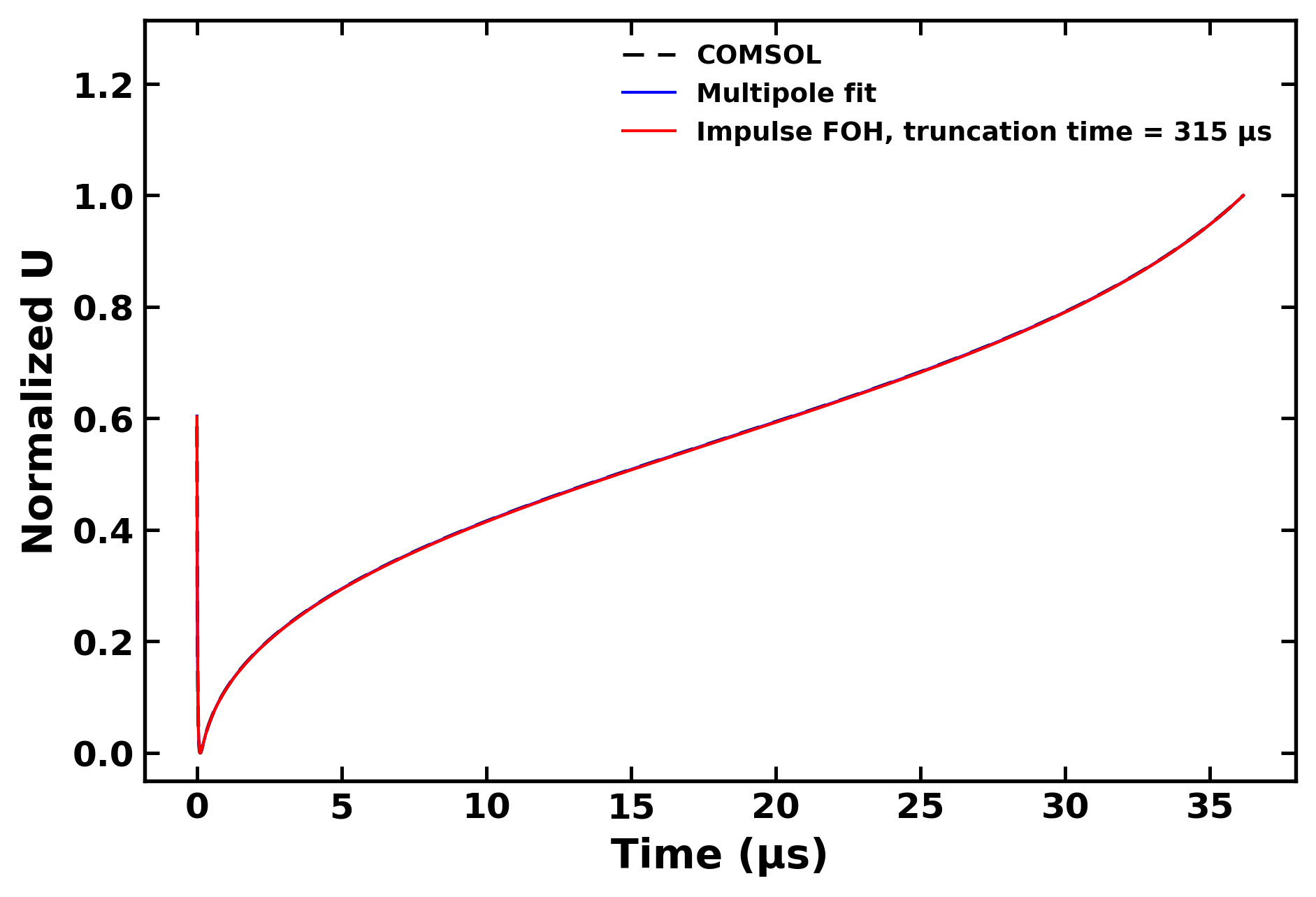}
        \caption{}
        \label{fig:norm_u_beginning}
    \end{subfigure}
    \hfill
    \begin{subfigure}{0.48\textwidth}
        \centering
        \includegraphics[width=\linewidth]{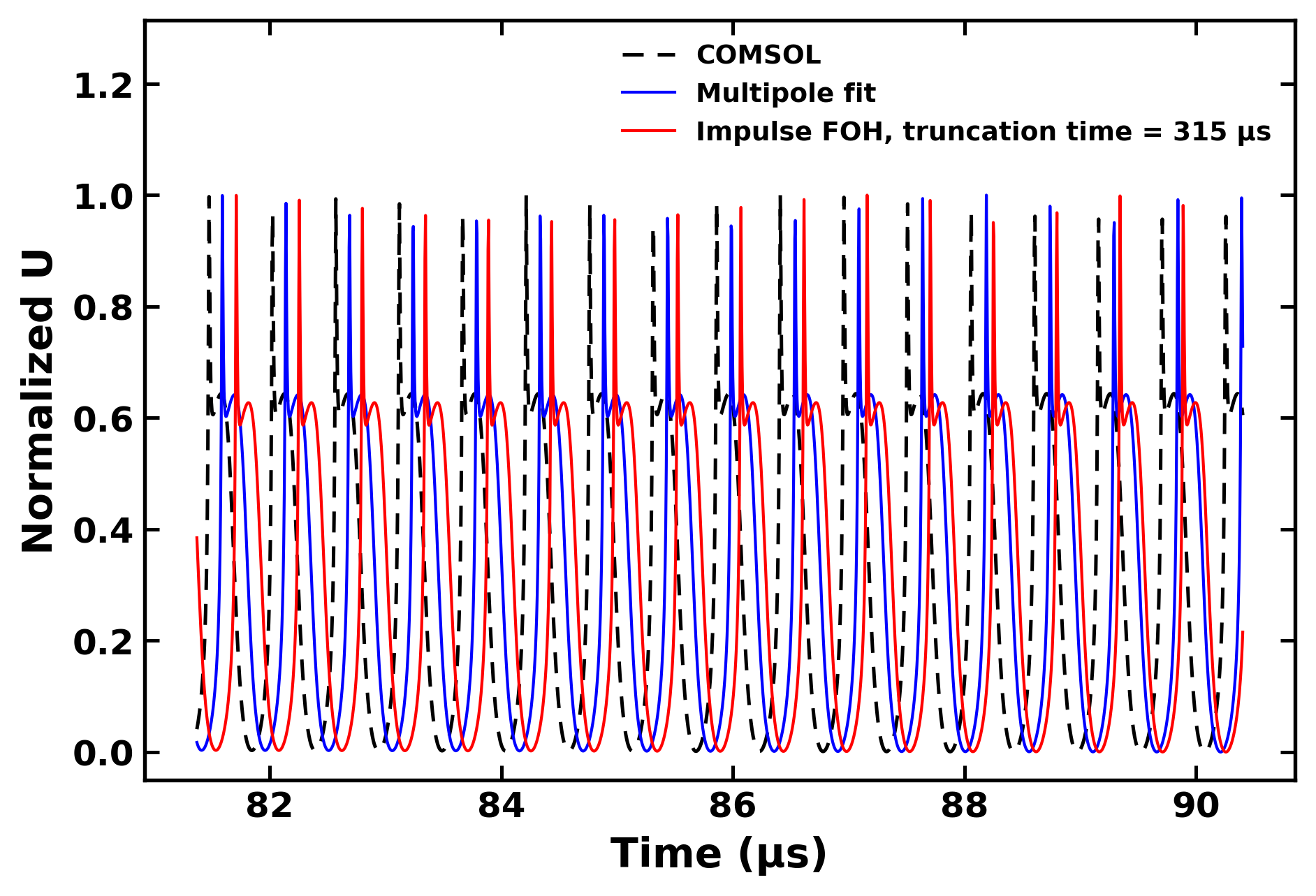}
        \caption{}
        \label{fig:norm_u_later}
    \end{subfigure}
    \hfill
    \begin{subfigure}{0.48\textwidth}
        \centering
        \includegraphics[width=\linewidth]{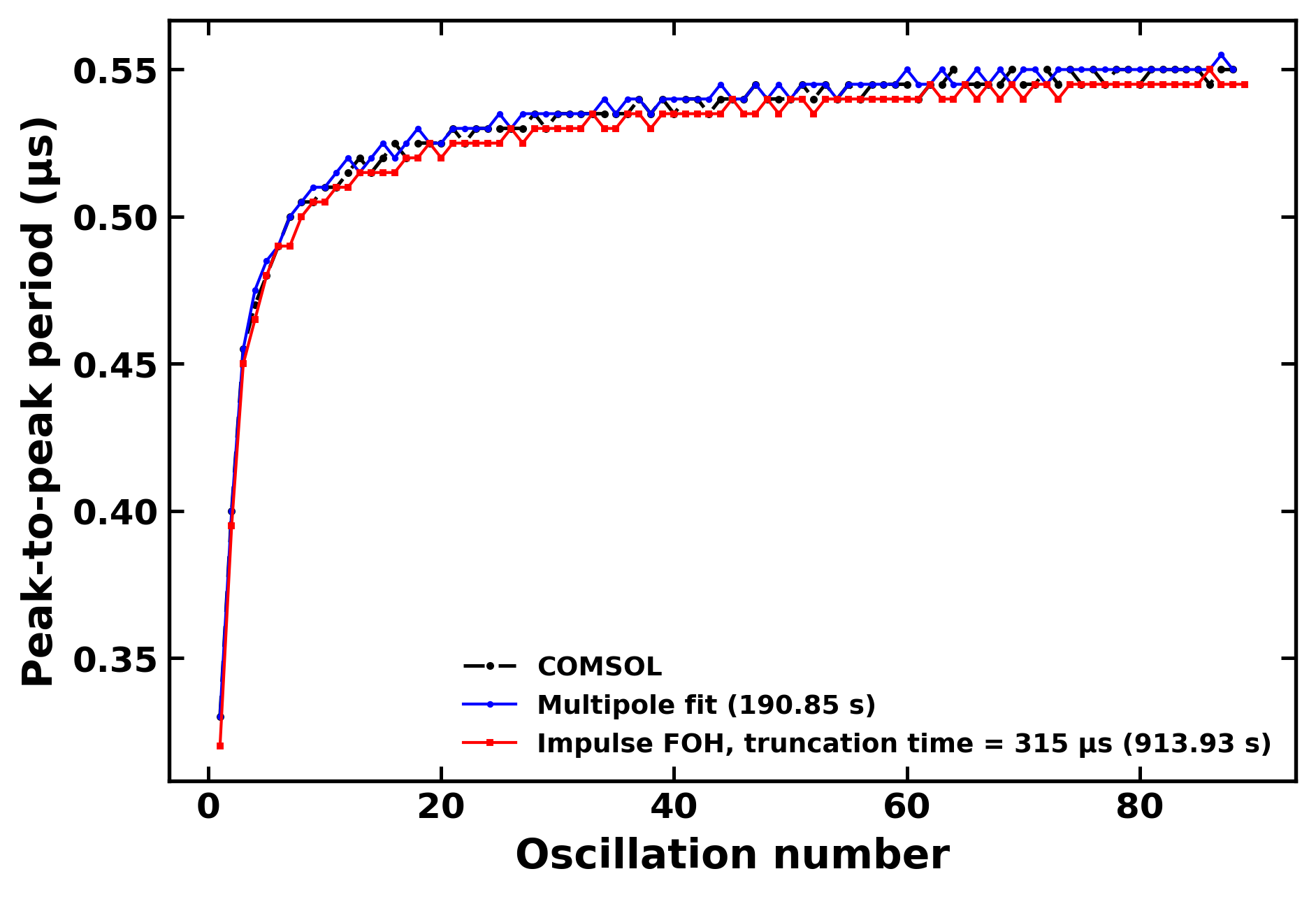}
        \caption{}
        \label{fig:benchmark_period}
    \end{subfigure}

    \caption{
    Comparison of the detailed reduced-order cavity simulations with the COMSOL
    time-domain reference. Panels (a) and (b) show the normalized intracavity
    energy $U=|a|^2$ over early and late time windows, respectively. Panel (c)
    compares the extracted maximum-to-maximum self-pulsation period.
    }
    \label{fig:detailed_cavity_simulation}
\end{figure}

For the parameters and dimensions provided by Borghi \textit{et
al.}~\cite{borghi2021modeling}, the reduced-order models were compared with
COMSOL time-domain simulations using identical optical and free-carrier
equations; the complete equations are provided in the Supplementary
Information. This comparison uses the detailed cavity model, including the
density-dependent Shockley--Read--Hall lifetime and separate free-carrier-
dispersion contributions. The only difference among the three simulations is
the thermal block: COMSOL solves Eq.~\eqref{eq:heat_equation_deltaT} directly,
whereas the reduced models use either the rational approximation or the
impulse-response representation of $H_{\mathrm{th}}(\omega)$. Unlike the
TE\(_1\) feed-forward test, the absorbed power is determined self-consistently
by the cavity state, so small differences in the thermal response can be
amplified by nonlinear feedback.

At early times, all three methods agree closely, as shown in
Fig.~\ref{fig:norm_u_beginning}. As self-pulsation persists, small differences
in oscillation period accumulate and progressively displace the time traces
[Fig.~\ref{fig:norm_u_later}]. This behavior is captured more clearly by the
period sequence in Fig.~\ref{fig:benchmark_period}. The reduced models therefore
reproduce the local self-pulsation dynamics, but exact long-time phase alignment
is sensitive to the thermal representation. Both approaches avoid repeated
solution of the heat equation, with the rational approximation providing the
shorter reduced-model runtime.

In both reduced models, $P_{\mathrm{abs}}$ depends on $T_{\mathrm{eff}}$
through the detuning, so the thermal response is coupled self-consistently rather
than evaluated as a post-processing step. Agreement with COMSOL was assessed
using $T_{\mathrm{eff}}(t)$, $N(t)$, $U(t)$, and the maximum-to-maximum
self-pulsation period. Because the optical and carrier equations are held fixed,
differences from COMSOL primarily quantify the error introduced by replacing
the full heat equation with either the rational approximation or the truncated
impulse-response representation.

Thus, the reduced-order models reproduce the local COMSOL waveform even when
detailed optical and carrier dynamics are retained. The main long-time
discrepancy is accumulated timing drift within the nonlinear limit cycle. For the simulations in Fig~\ref{fig:benchmark_period} while the COMSOL simulation takes a few days, the rational approximation takes 190.85s, and the impulse method takes 913s.

To separate this nonlinear sensitivity from the basic accuracy of the thermal
replacement, I also simulate the simplified cavity dynamics described in the
Supplementary Information for several input powers, simulation windows, and
modulation frequencies. These cases use a constant carrier lifetime and one
effective linear free-carrier-dispersion term,
\[
\tau_{\mathrm{fc}}=40~\mathrm{ns},
\qquad
\sigma_{\mathrm{eff}}=-9.5\times10^{-27}~\mathrm{m^3}.
\]
Their input powers and cold-cavity detunings differ from those of the detailed
comparison in Fig.~\ref{fig:detailed_cavity_simulation}; absolute oscillation
periods across these operating conditions should therefore not be compared
directly. The corresponding temperature errors are listed in
Table~\ref{tab:temperature_rmse_summary}.
\begin{table}[htbp]
\centering
\caption{Percentage temperature RMSE for different optical input and modulation
conditions. LC denotes a limit-cycle case, C a constant-input case without
self-pulsation, and M a modulated-input case.}
\label{tab:temperature_rmse_summary}
\begin{tabular}{l c l c}
\hline
Input condition & $\mathrm{RMSE}_{T}$ (\%) & Input condition & $\mathrm{RMSE}_{T}$ (\%) \\
\hline
$12~\mathrm{dBm}$ (LC) & 0.211 & $12~\mathrm{dBm}$ (LC), $274~\mu\mathrm{s}$ window & 2.184 \\
$11~\mathrm{dBm}$ (LC) & 0.289 & $10~\mathrm{dBm}$ (LC) & 0.548 \\
$5~\mathrm{dBm}$ (C) & 0.003 & $4~\mathrm{dBm}$ (C) & 0.003 \\
$2~\mathrm{dBm}\,\pm100\%$ at $1~\mathrm{GHz}$ (M) & 1.900 & $4~\mathrm{dBm}\,\pm12.5\%$ at $100~\mathrm{MHz}$ (M) & 0.886 \\
$4~\mathrm{dBm}\,\pm12.5\%$ at $100~\mathrm{kHz}$ (M) & 0.538 & $4~\mathrm{dBm}\,\pm12.5\%$ at $100~\mathrm{Hz}$ (M) & 0.438 \\
\hline
\end{tabular}
\end{table}

For these simplified cases, the rational approximation agrees closely with
COMSOL over the tested input and modulation conditions.

\section{Sensitivity to TTF}
\label{sec:sensitivity_ttf}

This sensitivity study uses the same simplified constant-lifetime,
single-FCD cavity model and the same rational-approximation procedure as the
Table~\ref{tab:temperature_rmse_summary} comparisons. Within each comparison,
the $12~\mathrm{dBm}$ optical input, cold-cavity detuning, and carrier
parameters are held fixed. Only the thermal transfer function is changed from
the centroid-sampled response $H(\mathbf r_c,\omega)$ to the globally weighted
effective response $H_{\mathrm{eff}}(\omega)$.

\begin{figure}[htbp]
    \centering

    \begin{subfigure}{0.48\textwidth}
        \centering
        \includegraphics[width=\linewidth]{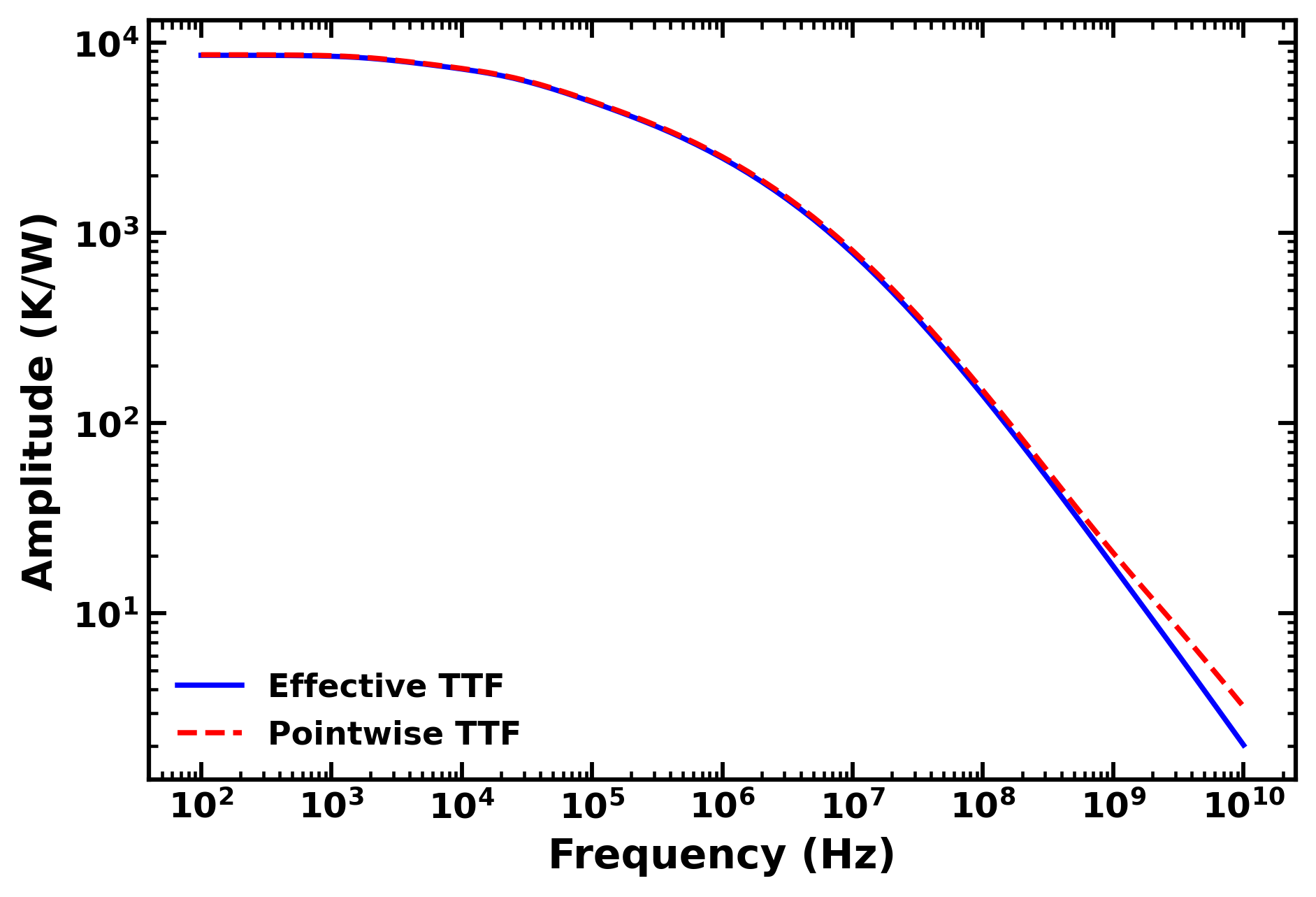}
        \caption{}
        \label{fig:subfig_a}
    \end{subfigure}
    \hfill
    \begin{subfigure}{0.48\textwidth}
        \centering
        \includegraphics[width=\linewidth]{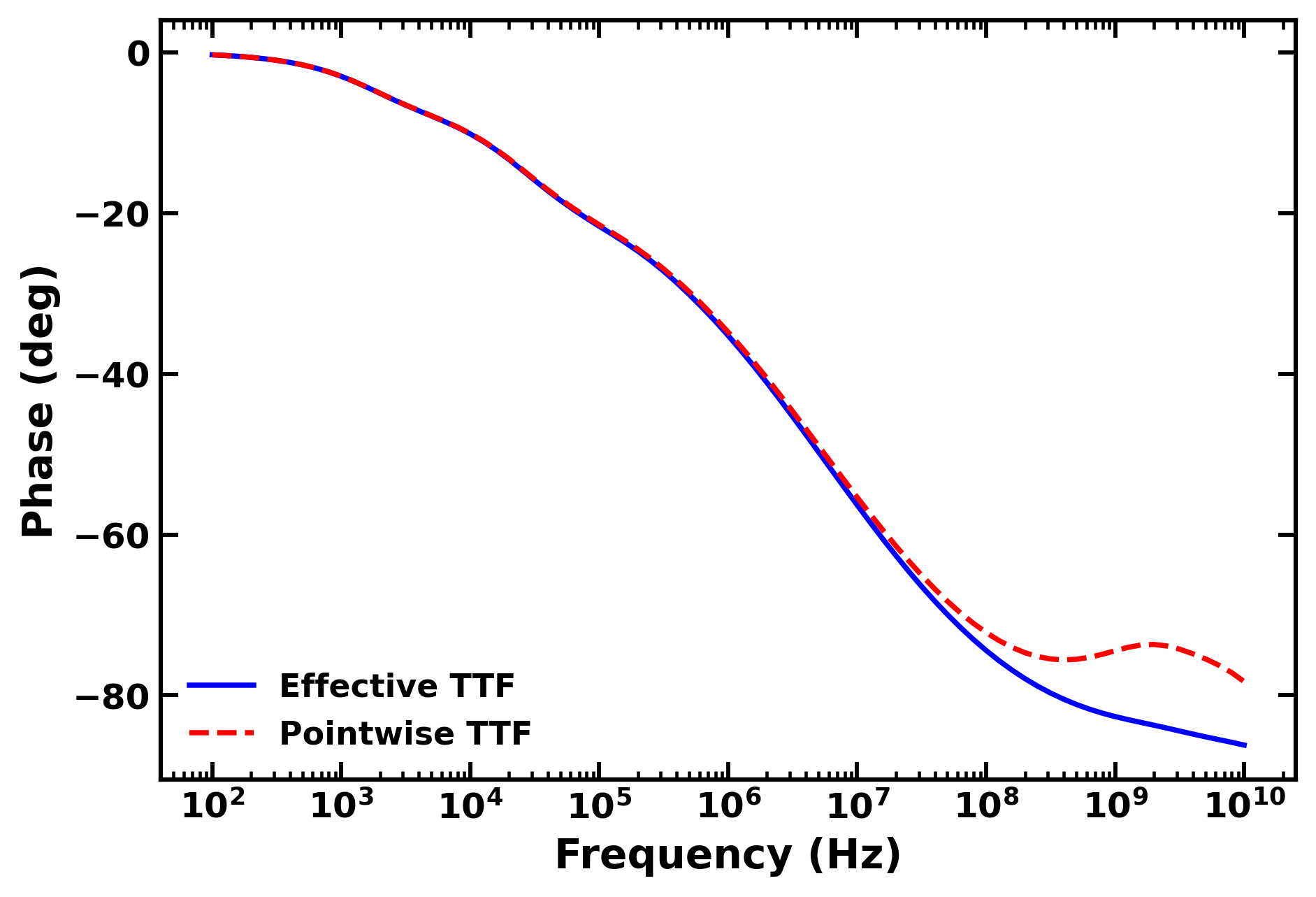}
        \caption{}
        \label{fig:subfig_b}
    \end{subfigure}

    \vspace{0.25cm}

    \begin{subfigure}{0.48\textwidth}
        \centering
        \includegraphics[width=\linewidth]{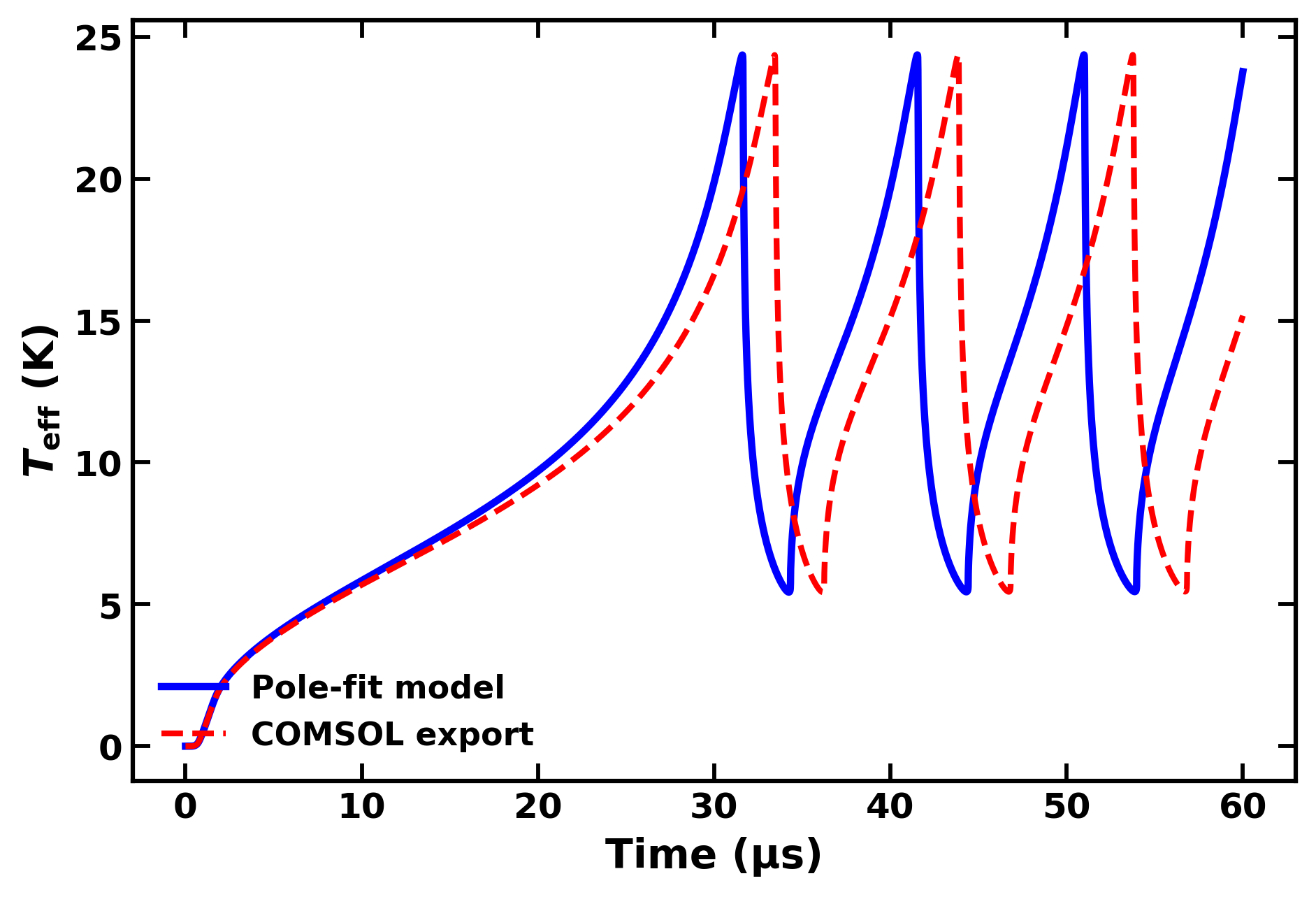}
        \caption{}
        \label{fig:subfig_c}
    \end{subfigure}
    \hfill
    \begin{subfigure}{0.48\textwidth}
        \centering
        \includegraphics[width=\linewidth]{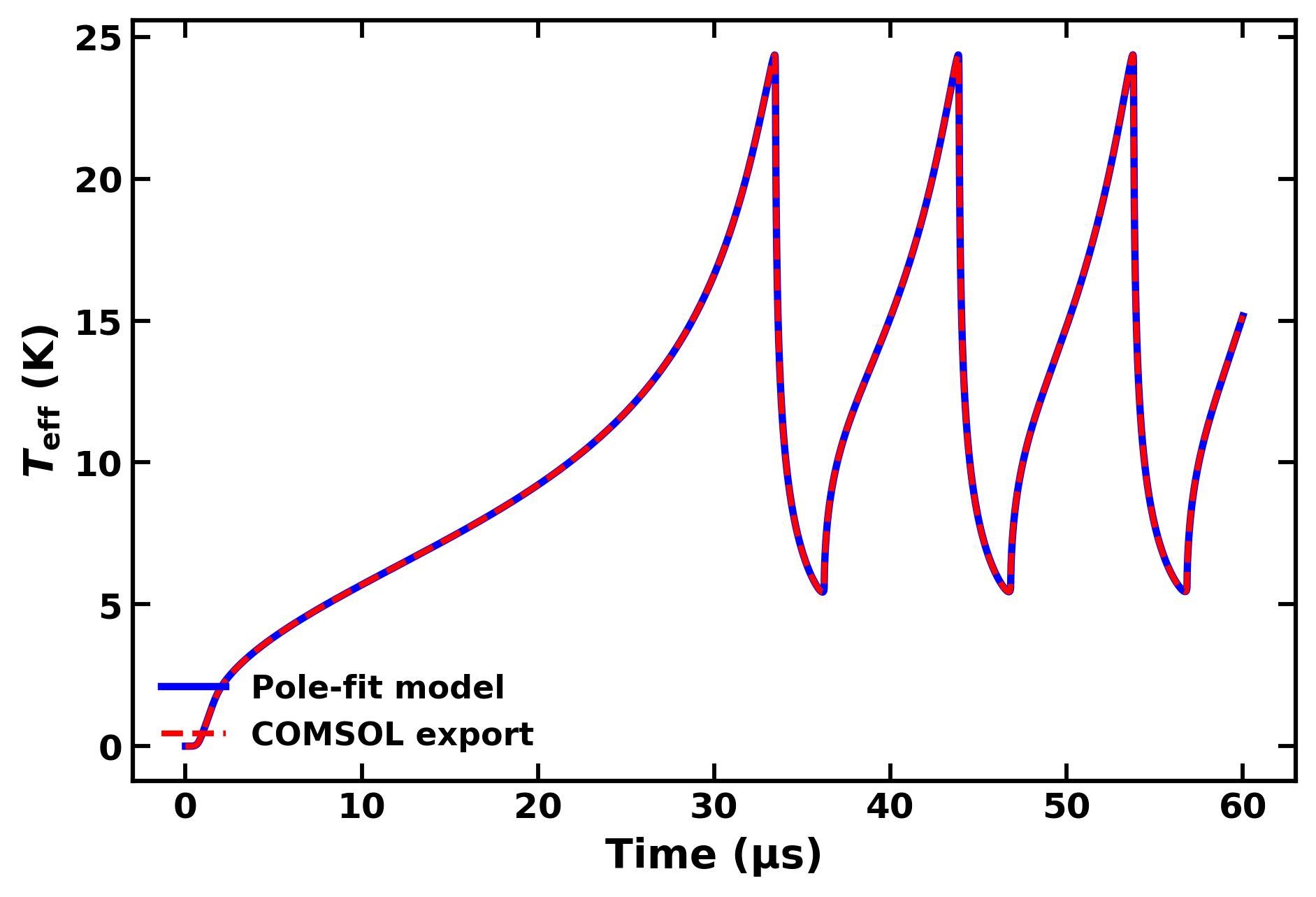}
        \caption{}
        \label{fig:subfig_d}
    \end{subfigure}

    \caption{
    Sensitivity of the nonlinear cavity dynamics to the spatial definition of
    the thermal transfer function. Panels (a) and (b) compare the amplitude and
    phase of the centroid response $H(\mathbf r_c,\omega)$ and the optically
    weighted effective response $H_{\mathrm{eff}}(\omega)$. Panels (c) and (d)
    compare the COMSOL temperature with the rational-approximation simulations
    obtained using the centroid and effective transfer functions, respectively.}
    \label{fig:two_by_two}
\end{figure}

The accumulated timing disparity in Fig.~\ref{fig:two_by_two} can be traced to
differences between the centroid-sampled and optically weighted thermal transfer
functions. In the nonlinear feedback regime, even modest high-frequency
differences alter the heating trajectory and the oscillation period
[Figs.~\ref{fig:subfig_c} and \ref{fig:subfig_d}]. By contrast, the TE$_1$
feed-forward study showed that the centroid response can approximate the global
effective temperature when the absorbed power is prescribed. For strongly
coupled nonlinear simulations, however, the transfer function must be derived
using the same optical weighting as the effective temperature in the cavity
model. The centroid approximation remains useful for point-probe calculations
or when its response has been shown to approximate the relevant weighted
temperature over the required bandwidth.

\section{Convergence of Impulse Response}
\label{sec:impulse_convergence}

An important practical parameter in the impulse-response method is the time
beyond which the impulse is neglected. Although the response is largest near
the origin, its tail contains the long-timescale thermal memory of the
structure. Truncating this tail reduces memory and computation time but can
alter nonlinear feedback. A $15~\mu\mathrm{s}$ kernel was sufficient for the
feed-forward TE$_1$ simulations in Sec.~\ref{sec:te1_heating}; the
self-pulsating cavity requires a longer window.

\begin{figure}[t]
    \centering

    \begin{minipage}[t]{0.48\textwidth}
        \centering
        \includegraphics[width=\linewidth]{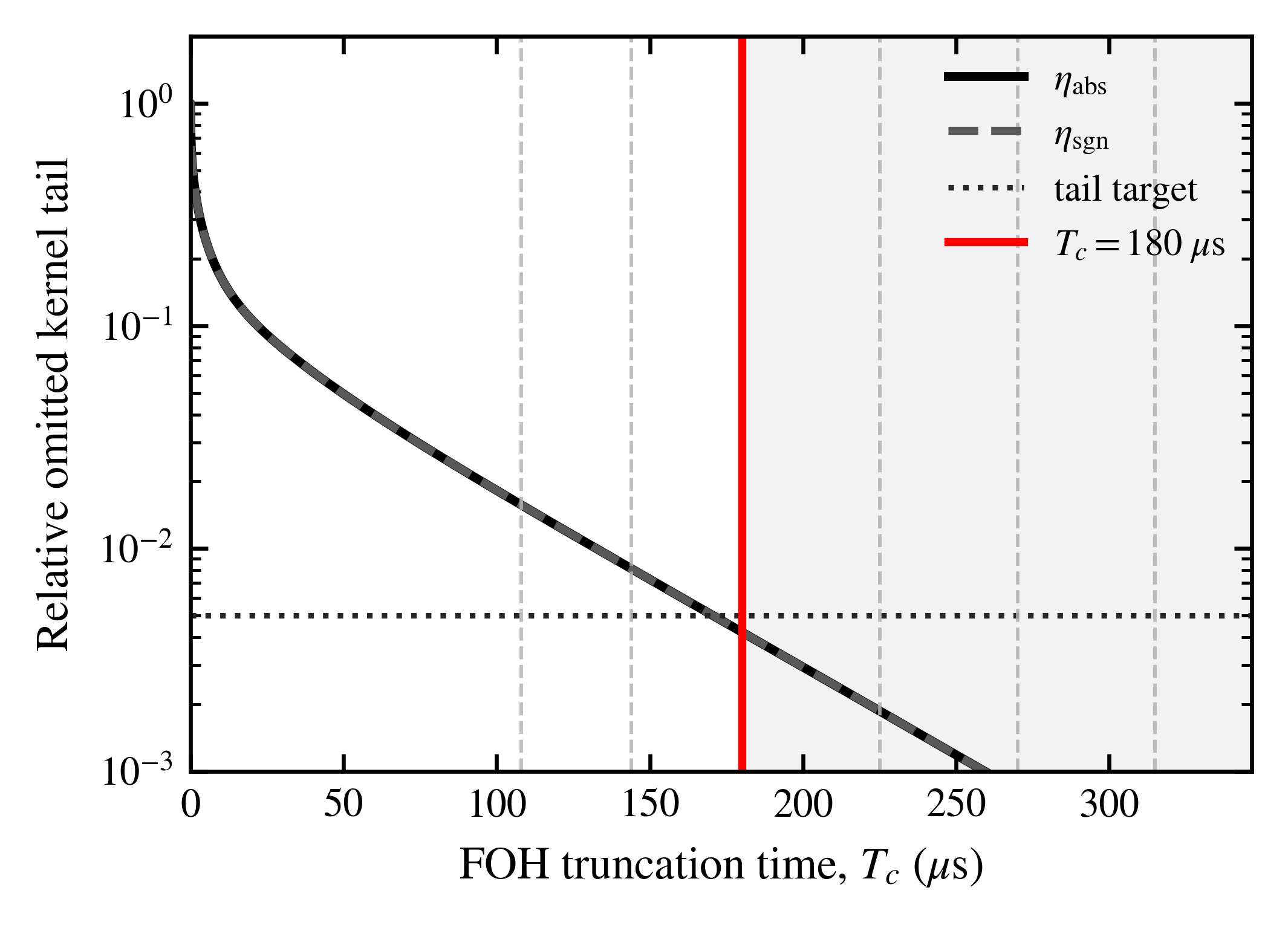}
        \textbf{(a)}
    \end{minipage}
    \hfill
    \begin{minipage}[t]{0.48\textwidth}
        \centering
        \includegraphics[width=\linewidth]{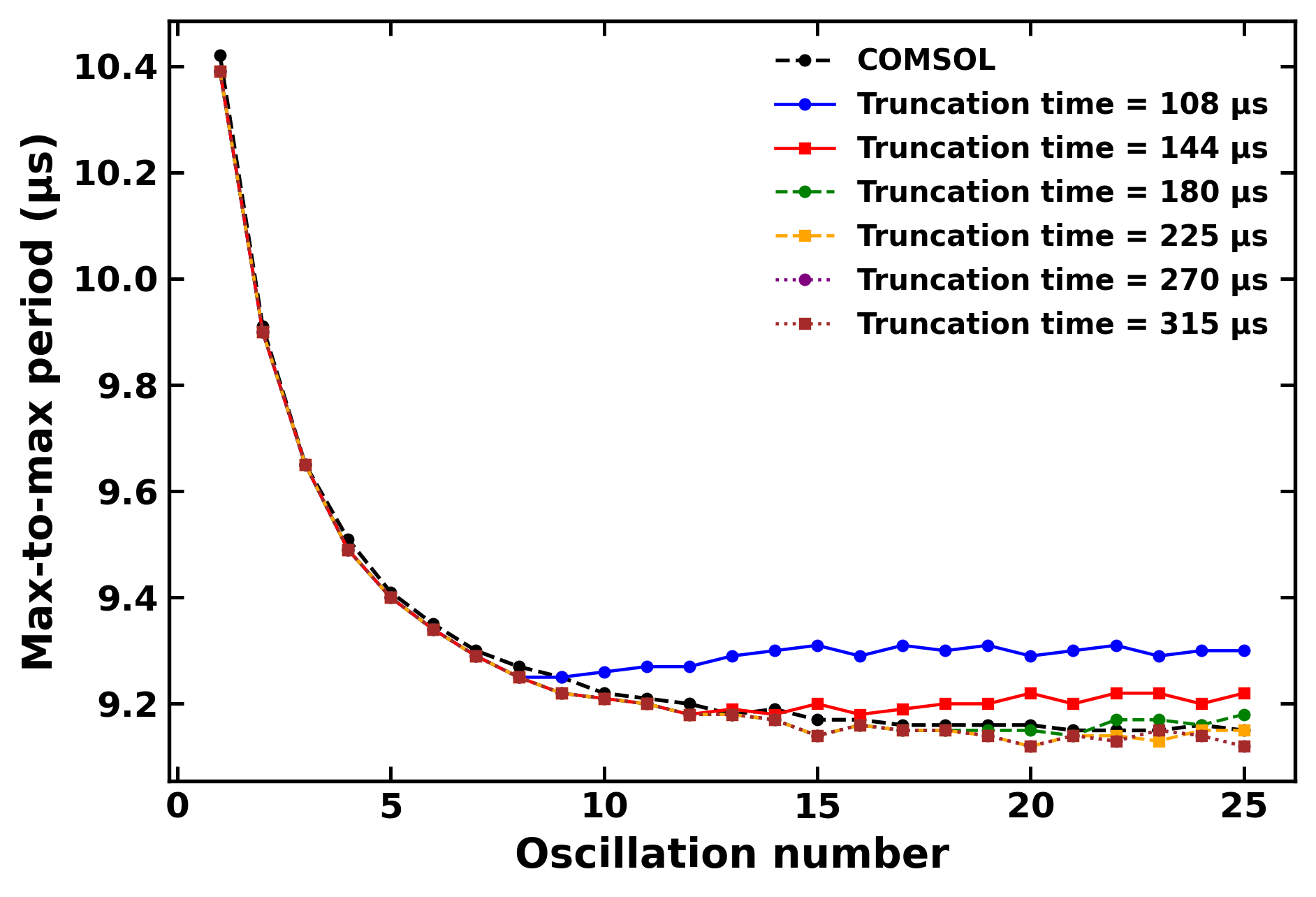}
        \textbf{(b)}
    \end{minipage}

    \caption{Convergence of the impulse-response window. (a) Identification of
    the critical truncation time (red) using a $5\times10^{-3}$ omitted-tail
    criterion; gray lines indicate the tested truncation times. (b)
    Maximum-to-maximum oscillation period versus oscillation number for the
    COMSOL reference and impulse-response simulations with different truncation
    times.}
    \label{fig:one_by_two}
\end{figure}

The convergence study in Fig.~\ref{fig:one_by_two} uses the simplified cavity
model with $\tau_{\mathrm{fc}}=40~\mathrm{ns}$ and
$\sigma_{\mathrm{eff}}=-9.5\times10^{-27}~\mathrm{m^3}$, together with the
exported $12~\mathrm{dBm}$ input trace, zero cold-cavity detuning, and a
$900~\mu\mathrm{s}$ simulation window. These conditions differ from the
detailed model and operating point used in Fig.~\ref{fig:detailed_cavity_simulation},
so their absolute self-pulsation periods are not expected to coincide.

Following Eq.~\ref{eq:discrete_conv}, the reference convolution is written as
\[
T_{\rm eff}[n]
=
\sum_{m=0}^{M_{\rm ref}-1} K[m] P_{\rm abs}[n-m].
\]
Here $M_{\rm ref}$ is the number of samples in a long reference kernel,
chosen to contain the relevant thermal memory. For a reference duration
$T_{\rm ref}$ and time step $\Delta t$,
$M_{\rm ref}=\lfloor T_{\rm ref}/\Delta t\rfloor$. A trial duration
$T_{\rm c}$ retains $M_{\rm c}=\lfloor T_{\rm c}/\Delta t\rfloor$ samples,
with indices $0\leq m<M_{\rm c}$; the omitted tail therefore begins at
$m=M_{\rm c}$. The kernel weight $K_m$ includes the appropriate time-step
factor. The absolute and signed omitted-tail metrics are
\begin{align}
\eta_{\rm abs}(T_{\rm c})
&=
\frac{
\sum_{m=M_{\rm c}}^{M_{\rm ref}-1} |K_m|
}{
\sum_{m=0}^{M_{\rm ref}-1} |K_m|
},
\label{eq:eta_abs}\\
\eta_{\rm sgn}(T_{\rm c})
&=
\frac{
\left|\sum_{m=M_{\rm c}}^{M_{\rm ref}-1} K_m\right|
}{
\left|\sum_{m=0}^{M_{\rm ref}-1} K_m\right|
}.
\label{eq:eta_sgn}
\end{align}
Here, $\eta_{\rm abs}$ measures the total magnitude of the discarded thermal
memory, whereas $\eta_{\rm sgn}$ measures its net signed contribution. The
absolute sum provides an upper bound because the discarded temperature is
\[
\Delta T_{\rm tail}[n]
=
\sum_{m=M_{\rm c}}^{M_{\rm ref}-1} K_m P_{\rm abs}[n-m].
\]

If $P_{\rm max}$ is an upper bound on the absorbed power over the time
history, then
\[
|\Delta T_{\rm tail}[n]|
\leq
P_{\rm max}
\sum_{m=M_{\rm c}}^{M_{\rm ref}-1} |K_m| .
\]
Therefore, $\eta_{\rm abs}$ provides a conservative measure of the remaining
kernel weight. The cutoff was chosen as the smallest $T_{\rm c}$ for which both
$\eta_{\rm abs}$ and $\eta_{\rm sgn}$ were below $5\times10^{-3}$, giving
$T_{\rm crit}=180~\mu\mathrm{s}$. Nonlinear simulations were then performed
with truncation times $\{0.6,0.8,1.0,1.25,1.5,1.75\}T_{\rm crit}$; the longest
window was therefore $315~\mu\mathrm{s}$. Because the cavity is self-pulsating,
convergence was evaluated using the maximum-to-maximum period as a function of
oscillation number.

The results show that a relatively long impulse window is required before the
period converges. This memory requirement is the principal computational
limitation of the impulse-response method. The rational approximation is faster
because it replaces the stored power history with a finite set of first-order
thermal states.

\section{Conclusion}

I have demonstrated that a harmonic thermal transfer function can replace the
time-dependent heat equation in nonlinear optical-microcavity simulations. Both
the rational approximation and impulse-response convolution retain the
distributed, multi-timescale thermal response and reproduce the principal
dynamics of the full finite-element model. The rational approximation provides
the greater computational reduction, whereas the impulse-response method avoids
fitting and follows directly from the sampled transfer function.

The comparisons also identify the conditions under which the reduction is
reliable. The transfer function must correspond to the same spatially weighted
temperature used in the optical model, particularly when nonlinear feedback
amplifies small thermal discrepancies. For impulse-response simulations, the
retained time window must also include enough of the thermal tail to preserve
the self-pulsation period. Subject to these conditions, transfer-function-based
thermal models provide a practical alternative to repeated transient
finite-element simulations. More generally, the same construction can be
applied to linear,
time-invariant PDEs whose source separates into fixed spatial and time-dependent
factors, as in Eq.~\eqref{eq:separable_source}.

\bibliography{sample}

@article{pernice2010time,
  title={Time-domain measurement of optical transport in silicon micro-ring resonators},
  author={Pernice, Wolfram HP and Li, Mo and Tang, Hong X},
  journal={Optics express},
  volume={18},
  number={17},
  pages={18438--18452},
  year={2010},
  publisher={Optical Society of America}
}

@article{johnson2006self,
  title={Self-induced optical modulation of the transmission through a high-Q silicon microdisk resonator},
  author={Johnson, Thomas J and Borselli, Matthew and Painter, Oskar},
  journal={Optics express},
  volume={14},
  number={2},
  pages={817--831},
  year={2006},
  publisher={Optical Society of America}
}

@article{borghi2021modeling,
  title={On the modeling of thermal and free carrier nonlinearities in silicon-on-insulator microring resonators},
  author={Borghi, Massimo and Bazzanella, Davide and Mancinelli, Mattia and Pavesi, Lorenzo},
  journal={Optics Express},
  volume={29},
  number={3},
  pages={4363--4377},
  year={2021},
  publisher={Optical Society of America}
}

@article{chandorkar2009multimode,
  title={Multimode thermoelastic dissipation},
  author={Chandorkar, Saurabh A and Candler, Robert N and Duwel, Amy and Melamud, Renata and Agarwal, Manu and Goodson, Kenneth E and Kenny, Thomas W},
  journal={Journal of applied physics},
  volume={105},
  number={4},
  year={2009},
  publisher={AIP Publishing}
}

@inproceedings{pavlov2023microresonator,
  title={Microresonator effective thermal parameters definition via thermal modes decomposition},
  author={Pavlov, Vladislav I and Kondratiev, Nikita M and Shitikov, Artem E and Lobanov, Valery E},
  booktitle={Photonics},
  volume={10},
  number={10},
  pages={1131},
  year={2023},
  organization={MDPI}
}

@article{straube2011thermal,
  title={Thermal wave propagation in thin films on substrate: The time-harmonic thermal transfer function},
  author={Straube, Hilmar and Breitenstein, Otwin and Wagner, Jan-Martin},
  journal={physica status solidi (b)},
  volume={248},
  number={9},
  pages={2128--2141},
  year={2011},
  publisher={Wiley Online Library}
}

@article{arendt2009weyl,
  title={Weyl's law: Spectral properties of the Laplacian in mathematics and physics},
  author={Arendt, Wolfgang and Nittka, Robin and Peter, Wolfgang and Steiner, Frank},
  journal={Mathematical analysis of evolution, information, and complexity},
  pages={1--71},
  year={2009},
  publisher={Wiley Online Library}
}

\end{document}


\title{Simulating optical microcavities by inverting the thermal transfer function: Supplementary Information}

\author{Sudipta Nayak}

\address{\authormark{1}Centre for Nano Science and Engineering, Indian Institute of Science, Bengaluru 560012, India}

\email{nayaks@anl.gov}


\section{Introduction}

This Supplementary Information provides the model definitions, numerical procedures, and
additional time-domain comparisons supporting the main text. The results use three related but
distinct dynamical configurations. First, the TE$_1$ feed-forward study tests whether a thermal
transfer function sampled at the optical centroid can approximate the globally evaluated effective
temperature. Second, the principal nonlinear-cavity comparison uses the detailed carrier model,
including density-dependent Shockley--Read--Hall recombination and separate free-carrier-
dispersion contributions. Third, the input-power sweep, the thermal-transfer-function sensitivity
study, and the impulse-kernel truncation study use the simplified Borghi-based cavity dynamics,
with a constant carrier lifetime and a single effective free-carrier-dispersion term.

These configurations also use different optical powers and cold-cavity detunings. Consequently,
their absolute temperatures and self-pulsation periods should not be compared as though they were
generated at a common operating point. The sections below identify the model and operating-case
assignment for each figure, provide the input definitions used for Table~2 of the main text, and
describe the convergence analysis for the truncated impulse-response model.

\section{Modal TE$_1$ Heating}

\begin{figure}[H]
\centering

\begin{subfigure}[t]{0.32\textwidth}
    \centering
    \includegraphics[width=\linewidth]{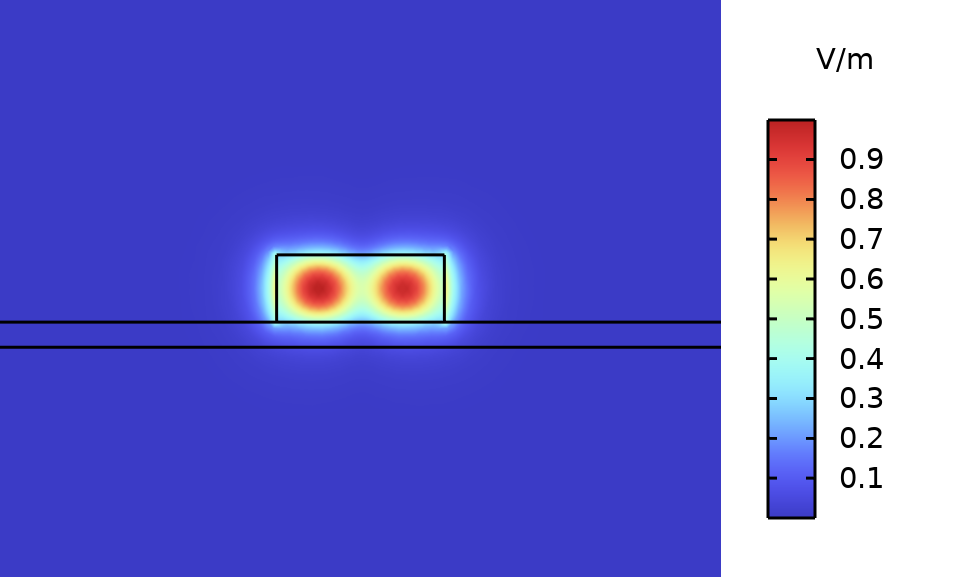}
    \caption{}
    \label{fig:supp_te1_mode_profile}
\end{subfigure}
\hfill
\begin{subfigure}[t]{0.32\textwidth}
    \centering
    \includegraphics[width=\linewidth]{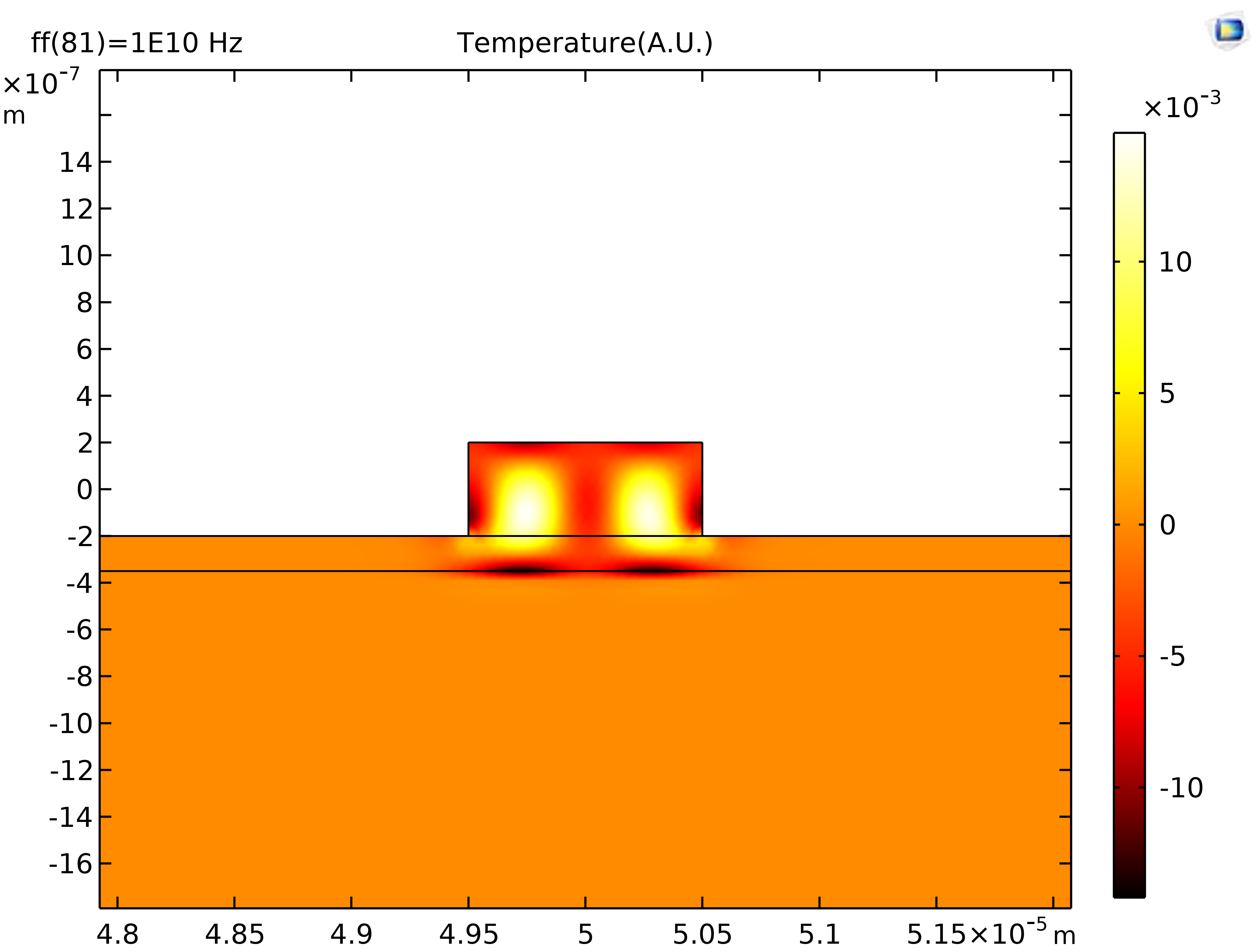}
    \caption{}
    \label{fig:supp_te1_ttf_real_10GHz}
\end{subfigure}
\hfill
\begin{subfigure}[t]{0.32\textwidth}
    \centering
    \includegraphics[width=\linewidth]{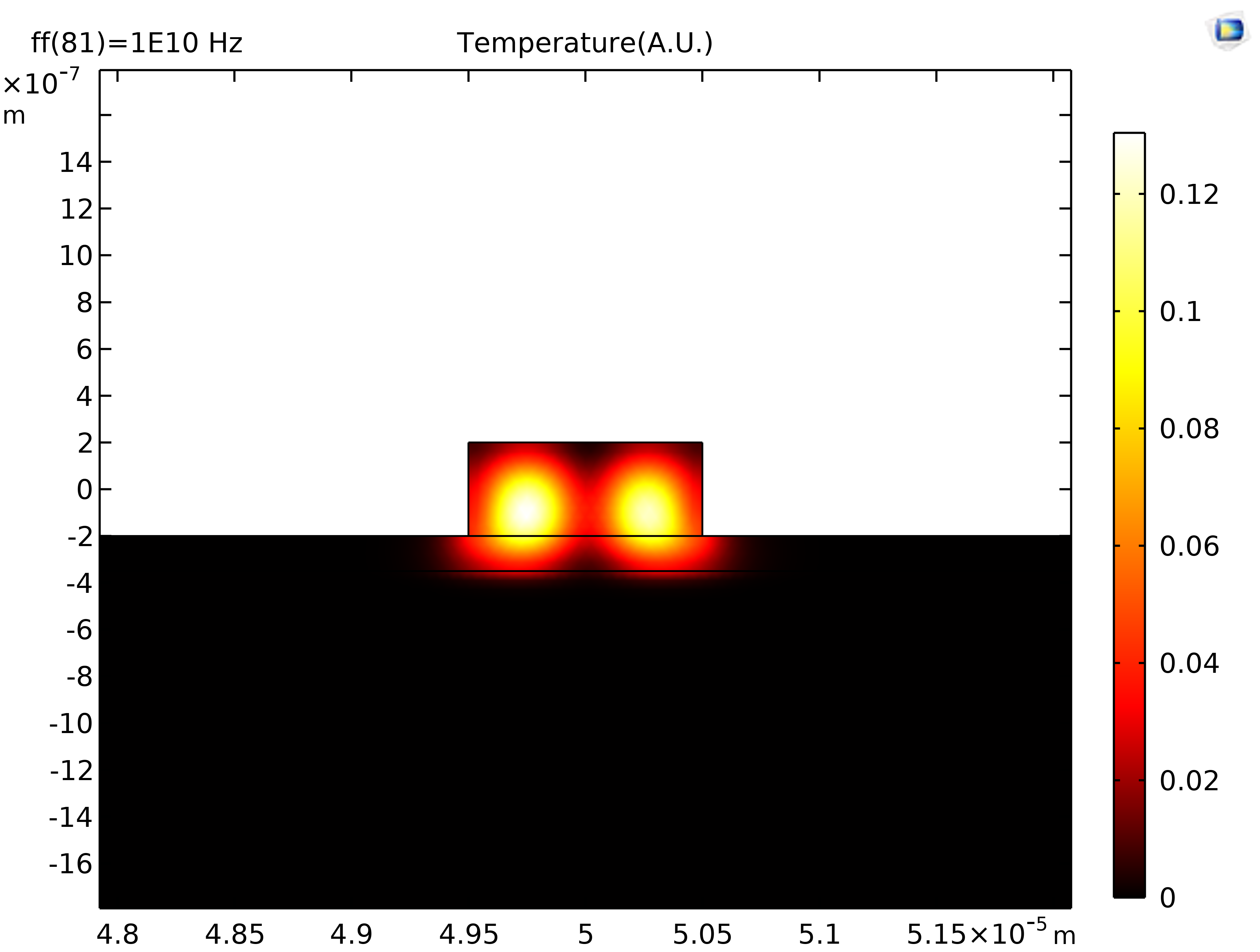}
    \caption{}
    \label{fig:supp_te1_ttf_imag_10GHz}
\end{subfigure}

\caption{
Spatial thermal-transfer-function fields for TE$_1$-mode heating.
(a) TE$_1$ optical mode profile used to define the normalized heat-source distribution
\(X_{\rm TE1}(\mathbf{r})\). 
(b) Real part and (c) imaginary part of the spatial thermal transfer function
\(H(\mathbf{r},\omega)\) at \(f=10~{\rm GHz}\). The finite spatial structure visible in the
TTF reflects the short thermal diffusion length at high modulation frequency.
}
\label{fig:supp_te1_ttf_spatial}
\end{figure}

The harmonic heat-equation solution used in the finite-element solver can be interpreted as the
spatial profile of the complex thermal transfer function. Following Eqs.~(1)--(5) of the main text,
a unit harmonic heat source with spatial profile \(X_{\rm TE1}(\mathbf{r})\) ($Q(\mathbf r,t)=X_{\mathrm{TE1}}(\mathbf r)\sin(\omega t)$) gives
\(\widetilde{T}(\mathbf{r},\omega)\), such that
\(\widetilde{T}(\mathbf{r},\omega)=H(\mathbf{r},\omega)\).
Thus, for every modulation frequency, the FEA solution is not a single scalar response but a
complex spatial TTF profile. Figure~\ref{fig:supp_te1_ttf_spatial} shows the TE$_1$ mode profile
used to construct \(X_{\rm TE1}(\mathbf{r})\), followed by the real and imaginary parts of
\(H(\mathbf{r},\omega)\) at \(10~{\rm GHz}\). At this high frequency, the reduced thermal diffusion
length allows spatial features of the optical heating profile and the material boundaries to remain
visible in both quadratures of the thermal response.

To reproduce the time-domain thermal dynamics, the spatial TTF was sampled at the centroid
\(\mathbf{r}_c\) defined in Eq.~(8) of the main text. The resulting scalar transfer function
\(H(\mathbf{r}_c,\omega)\) was inverted to obtain a causal impulse response. The temperature was
then computed using the discrete convolution form of Eq.~(13),
\begin{equation}
T(\mathbf{r}_c,n)
=
\sum_{m=0}^{M}K[m]P_{\rm abs}[n-m],
\end{equation}
where \(K[m]\) includes the impulse-response sample and the time-step factor. The inverse
Fourier transform was performed directly on the native frequency grid of the sampled TTF, with conjugate
completion used to obtain a real time-domain kernel. No adjustable fitting parameter was used in
the time-domain comparison. The TE$_1$ kernel was truncated at \(15~\mu{\rm s}\), chosen from
convergence of the impulse tail.
The full transient reference was generated by solving the time-dependent heat equation with the
same TE$_1$-derived source \(X_{\rm TE1}(\mathbf{r})\sin(\omega t)\). Its output was the globally
evaluated effective temperature obtained using the optical weighting defined in the main text.
The convolution result, in contrast, was generated from the point-sampled centroid TTF.
Therefore, this comparison tests the combined centroid approximation and impulse-response
inversion, rather than the inversion error alone.

\begin{figure}[t]
\centering

\begin{subfigure}[t]{0.48\textwidth}
    \centering
    \includegraphics[width=\linewidth]{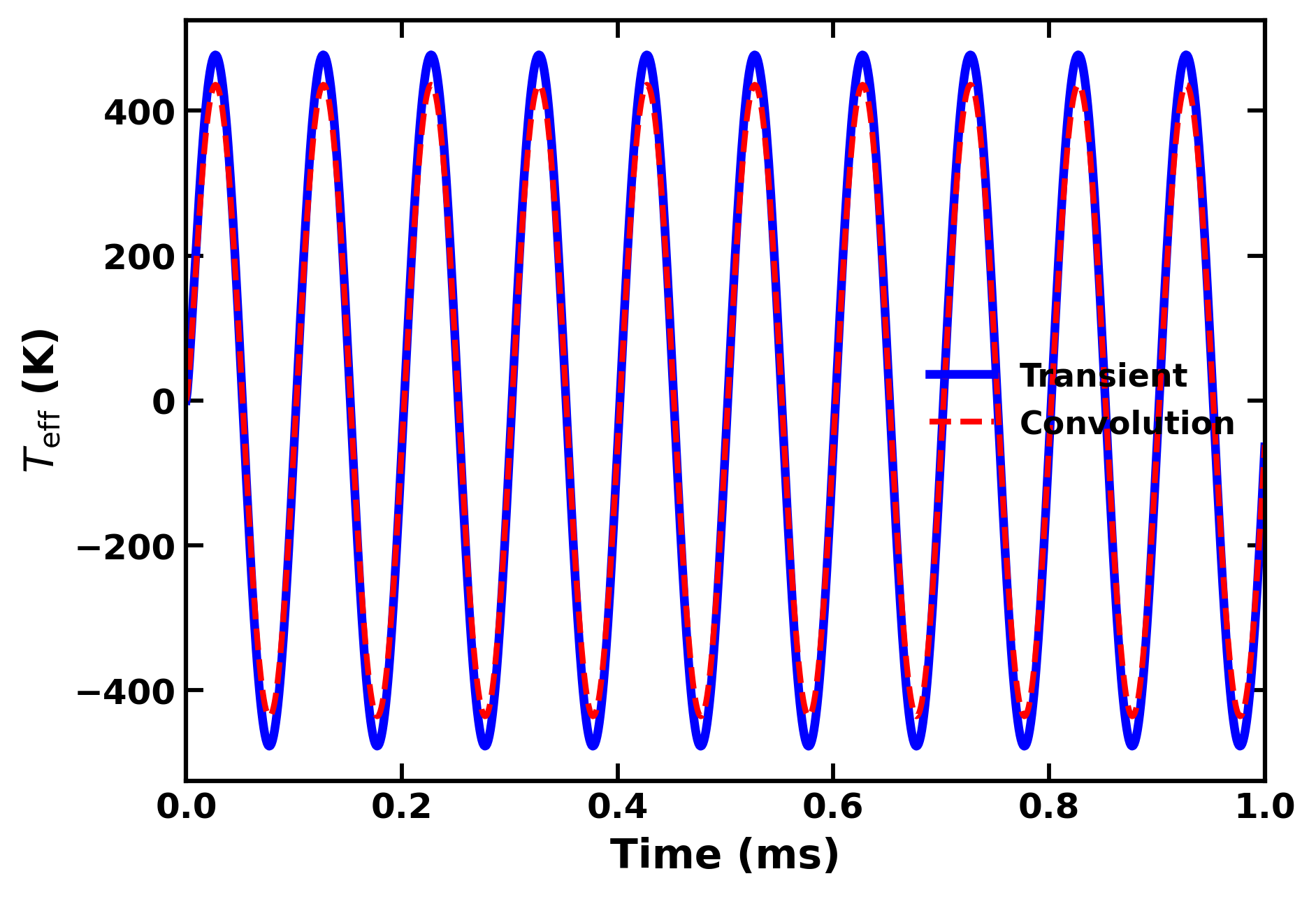}
    \caption{\(10~{\rm kHz}\), \(\mathrm{RMSE}_T=3.093\%\)}
\end{subfigure}
\hfill
\begin{subfigure}[t]{0.48\textwidth}
    \centering
    \includegraphics[width=\linewidth]{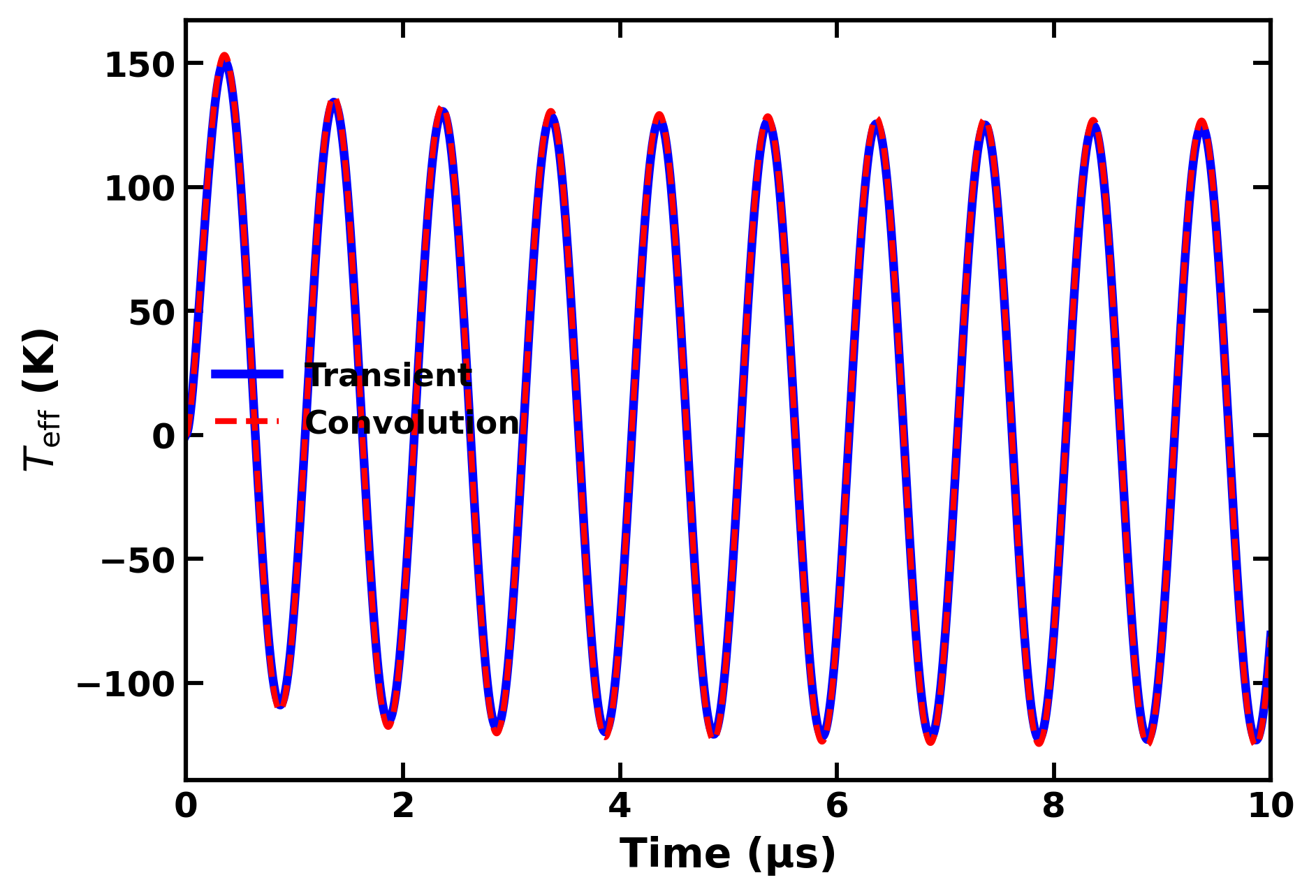}
    \caption{\(1~{\rm MHz}\), \(\mathrm{RMSE}_T=0.223\%\)}
\end{subfigure}

\vspace{0.6em}

\begin{subfigure}[t]{0.48\textwidth}
    \centering
    \includegraphics[width=\linewidth]{figures/Journal_TE1/100MHz.png}
    \caption{\(100~{\rm MHz}\), \(\mathrm{RMSE}_T=3.377\%\)}
\end{subfigure}
\hfill
\begin{subfigure}[t]{0.48\textwidth}
    \centering
    \includegraphics[width=\linewidth]{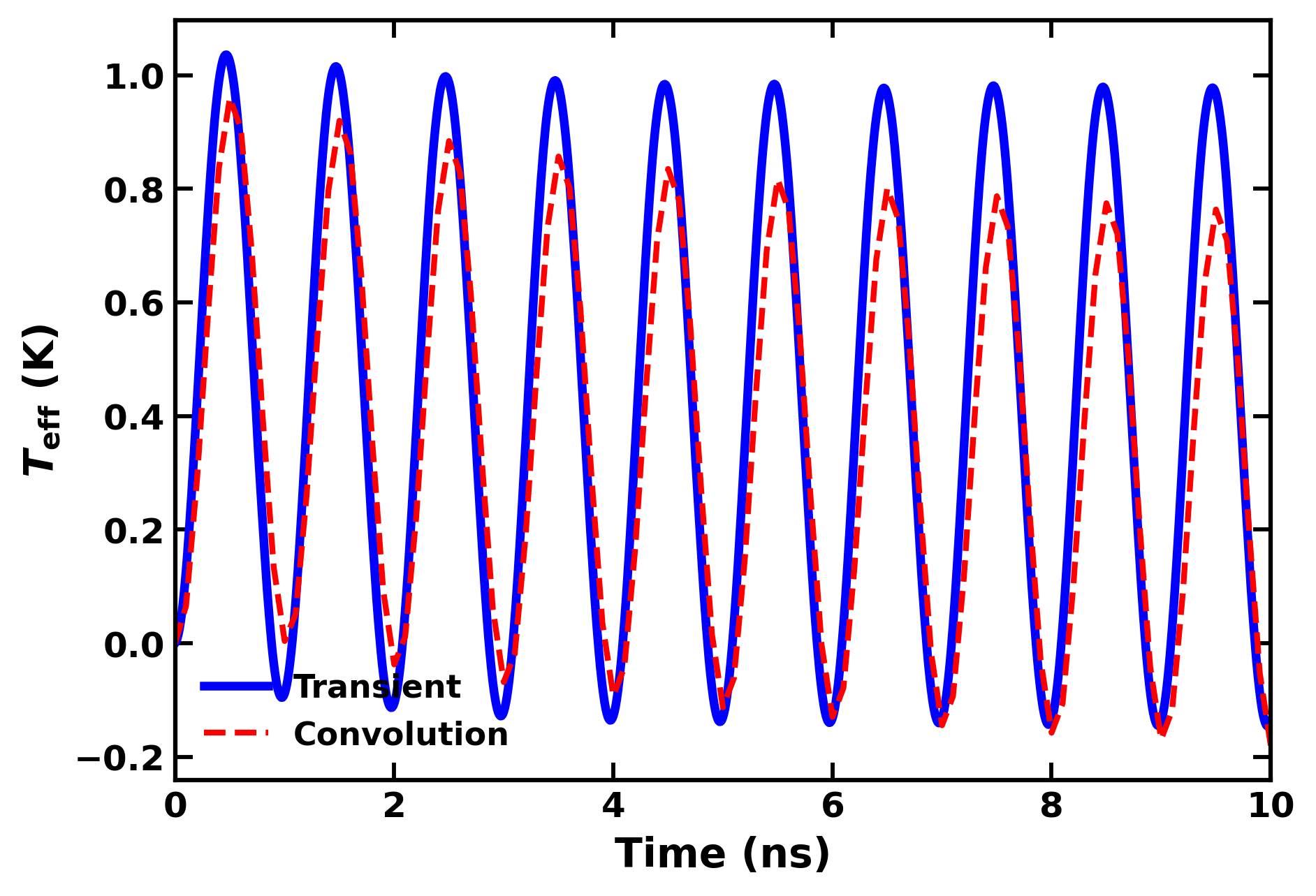}
    \caption{\(1~{\rm GHz}\), \(\mathrm{RMSE}_T=12.108\%\)}
\end{subfigure}

\caption{
Time-domain traces used to generate the TE$_1$ RMSE values reported in Table~1 of the main text.
Each panel compares the globally evaluated effective temperature from the full transient FEA
simulation with the temperature obtained by convolution using the impulse response derived from
the centroid-sampled thermal transfer function. The deviation therefore includes the effect of
replacing the global optical weighting by a pointwise centroid approximation.
}
\label{fig:supp_te1_timeseries_all}
\end{figure}

The time traces used to generate the RMSE values in Table~1 of the main text are shown in
Fig.~\ref{fig:supp_te1_timeseries_all}. The percentage temperature RMSE was evaluated as
\begin{equation}
{\rm RMSE}_{T,\%}
=
100
\frac{
\sqrt{
N^{-1}
\sum_{i=1}^{N}
\left(
T_{{\rm TTF},i}-T_{{\rm FEA},i}
\right)^2
}
}{
\sqrt{
N^{-1}
\sum_{i=1}^{N}
T_{{\rm FEA},i}^{2}
}
}.
\end{equation}
The agreement remains good up to \(100~{\rm MHz}\). The increased deviation at \(1~{\rm GHz}\)
is consistent with the stronger spatial structure of the high-frequency TTF and the use of a
centroid-sampled scalar response.

\section{Cavity Simulation}
\label{sec:supp_reduced_model}

This section describes the reduced-order model used to simulate the nonlinear
microring response in Fig.~2 of the main text. This is the detailed cavity
model: it includes density-dependent carrier recombination and separate
free-carrier-dispersion contributions. The optical power and cold-cavity
detuning are those of the corresponding detailed COMSOL simulation and differ
from the operating conditions used for the simplified comparisons in
Table~2 and Figs.~3--4 of the main text. The model combines a fast optical cavity field, a
free-carrier-rate equation, and either a rational thermal-transfer-function
model or an impulse-response thermal model. The optical field is assumed to
reach steady state much faster than the carrier and thermal variables. Therefore,
in the fast-field limit, the complex cavity field
\[
a(t)=a_r(t)+i a_i(t)
\]
is obtained algebraically at each time step rather than by integrating a
separate optical-field equation. The real and imaginary field quadratures,
\(a_r\) and \(a_i\), satisfy
\[
\gamma a_r+\Delta a_i-\sqrt{2\gamma_e}\,p_{\mathrm{in}} = 0,
\]
\[
-\Delta a_r+\gamma a_i = 0.
\]
Here \(p_{\mathrm{in}}\) is the input optical wave amplitude in the coupled-mode
normalization used in the simulations, \(\gamma_e\) is the coupling rate to each
bus waveguide, \(\gamma\) is the total cavity field decay rate, and
\(\Delta\) is the instantaneous detuning between the laser and the thermally and
carrier-shifted cavity resonance. The intracavity energy-like variable is
defined as
\[
U(t)=|a(t)|^2=a_r^2(t)+a_i^2(t).
\]

The total cavity loss rate includes coupling, intrinsic loss, free-carrier
absorption, and two-photon absorption:
\[
\gamma =
2\gamma_e+\gamma_i+\eta_{\mathrm{FCA}}N+\eta_{\mathrm{TPA}}U .
\]
Here \(\gamma_i\) is the intrinsic cavity loss rate, \(N(t)\) is the effective
free-carrier density, \(\eta_{\mathrm{FCA}}\) is the free-carrier-absorption
loss coefficient, and \(\eta_{\mathrm{TPA}}\) is the two-photon-absorption loss
coefficient. The factor \(2\gamma_e\) accounts for the two bus-waveguide
coupling channels.

The absorbed optical power that drives the thermal model is written as
\[
P_{\mathrm{abs}}
=
2\left(
f_i\gamma_i+\eta_{\mathrm{FCA}}N+\eta_{\mathrm{TPA}}U
\right)U .
\]
The coefficient \(f_i\) specifies the fraction of intrinsic optical loss that is
converted into heat. In the simulations reported here, \(f_i=1\), so all
intrinsic loss is assumed to contribute to local heating. The prefactor of 2
converts the field decay-rate convention into an energy-loss rate.

The optical detuning includes the cold-cavity detuning, thermo-optic feedback,
and free-carrier-dispersion feedback:
\[
\Delta
=
2\pi\Delta\nu_0
+
\omega_0
\left[
-\frac{1}{n_{\mathrm{Si}}}\frac{dn}{dT}T_{\mathrm{eff}}
-
\frac{\sigma_1N+\sigma_2N^{0.8}}{n_{\mathrm{Si}}}
\right].
\]
Here \(\Delta\nu_0\) is the laser detuning from the cold-cavity resonance in
Hz, \(\omega_0\) is the cold-cavity angular resonance frequency,
\(n_{\mathrm{Si}}\) is the refractive index of silicon, \(dn/dT\) is the
thermo-optic coefficient, and \(T_{\mathrm{eff}}(t)\) is the effective
temperature rise sensed by the optical mode. The coefficients \(\sigma_1\) and
\(\sigma_2\) describe the linear and sublinear free-carrier-dispersion
contributions, respectively.

\subsection{Free-carrier rate equation and SRH lifetime model}

The free-carrier density evolves according to
\[
\frac{dN}{dt}
=
g_{\mathrm{TPA}}U^2
-
\frac{N}{\tau_{\mathrm{fc}}(N)} ,
\]
where \(g_{\mathrm{TPA}}\) is the two-photon-absorption carrier-generation
coefficient and \(\tau_{\mathrm{fc}}(N)\) is the density-dependent effective
free-carrier lifetime. The lifetime was modeled using the Shockley--Read--Hall
expression used by Borghi \textit{et al.} for silicon-on-insulator microring
resonators~\cite{borghi2021modeling}. In that model, recombination is assumed to be
dominated by trap-assisted recombination at the Si/SiO\(_2\) interface, and the
instantaneous free-carrier lifetime is written as
\[
\tau_{\mathrm{fc}}(N)
=
\frac{N}{R(N)}
=
\tau_0
\frac{1+a_{\mathrm{SRH}}N}{1+b_{\mathrm{SRH}}N}.
\]
Here \(R(N)\) is the recombination rate, \(\tau_0\) is the low-injection
free-carrier lifetime, and \(a_{\mathrm{SRH}}\) and \(b_{\mathrm{SRH}}\) are
coefficients determined by the equilibrium carrier densities and by the electron
and hole trap-assisted recombination lifetimes. Following the notation of
Borghi \textit{et al.}, these coefficients are
\[
a_{\mathrm{SRH}}
=
\frac{\tau_p+\tau_n}
{\tau_p(n_1+n_0)+\tau_n(p_1+p_0)},
\]
\[
b_{\mathrm{SRH}}
=
\frac{1}{n_0+p_0},
\]
and
\[
\tau_0
=
\tau_p\frac{n_0+n_1}{n_0+p_0}
+
\tau_n\frac{p_0+p_1}{n_0+p_0}.
\]
In these expressions, \(\tau_n\) and \(\tau_p\) are the characteristic electron
and hole recombination lifetimes, \(n_0\) and \(p_0\) are the equilibrium
electron and hole densities, and \(n_1\) and \(p_1\) are the carrier densities
obtained when the Fermi level coincides with the trap energy level \(E_t\). They
are defined as
\[
n_1
=
n_i
\exp\left(
\frac{E_t-E_i}{k_B T_0}
\right),
\]
\[
p_1
=
n_i
\exp\left(
-\frac{E_t-E_i}{k_B T_0}
\right),
\]
where \(n_i\) is the intrinsic carrier concentration, \(E_i\) is the intrinsic
Fermi level, \(k_B\) is Boltzmann's constant, and \(T_0\) is the equilibrium
temperature. For a p-type silicon waveguide, the equilibrium electron density
was taken as
\[
n_0=\frac{n_i^2}{p_0}.
\]
This density-dependent lifetime model allows the recombination rate to vary
with the instantaneous carrier population, instead of assuming a constant
carrier lifetime.

The main constants used in the nonlinear cavity simulations are listed in
Table~\ref{tab:cavity_constants}. Unless otherwise stated, the same constants
were used in the COMSOL-matched reduced model and in the
thermal-transfer-function simulations.

\begin{table}[htbp]
\centering
\caption{Model constants used in the nonlinear cavity simulations.}
\label{tab:cavity_constants}
\small
\begin{tabular}{lll}
\hline
Parameter & Value & Meaning \\
\hline
$\omega_0/2\pi$ & $192.88~\mathrm{THz}$ & Optical resonance frequency \\
$\gamma_e$ & $43.0\times10^{9}~\mathrm{s^{-1}}$ & Coupling rate per bus \\
$\gamma_i$ & $5.5\times10^{9}~\mathrm{s^{-1}}$ & Intrinsic loss rate \\
$n_{\mathrm{Si}}$ & $3.485$ & Silicon refractive index \\
$n_g$ & $4.4$ & Group index \\
$dn/dT$ & $1.86\times10^{-4}~\mathrm{K^{-1}}$ & Thermo-optic coefficient \\
$\sigma_1$ & $-1.07\times10^{-27}~\mathrm{m^3}$ & Linear FCD coefficient \\
$\sigma_2$ & $-1.63\times10^{-22}~\mathrm{m^{2.4}}$ & Sublinear FCD coefficient \\
$\sigma_{\mathrm{FCA}}$ & $1.5\times10^{-21}~\mathrm{m^2}$ & FCA cross-section \\
$\eta_{\mathrm{FCA}}$ & $c_0\sigma_{\mathrm{FCA}}/n_g$ & FCA loss coefficient \\
$\eta_{\mathrm{TPA}}$ & $5.56\times10^{21}$ & TPA loss coefficient \\
$g_{\mathrm{TPA}}$ & $9.88\times10^{57}$ & TPA carrier-generation coefficient \\
$\tau_n$ & $35.0~\mathrm{ns}$ & SRH electron lifetime \\
$\tau_p$ & $22.5~\mathrm{ns}$ & SRH hole lifetime \\
$N(0)$ & $10^8~\mathrm{m^{-3}}$ & Initial carrier density \\
\hline
\end{tabular}
\end{table}

The SRH parameters used to evaluate
\(\tau_{\mathrm{fc}}(N)\) are listed in
Table~\ref{tab:srh_parameters}. Carrier densities reported in
\(\mathrm{cm^{-3}}\) were converted to \(\mathrm{m^{-3}}\) in the numerical
implementation so that they are consistent with the carrier-density variable
\(N\).

\begin{table}[htbp]
\centering
\caption{Parameters used in the Shockley--Read--Hall free-carrier lifetime model.}
\label{tab:srh_parameters}
\small
\begin{tabular}{lll}
\hline
Parameter & Value & Meaning \\
\hline
$\tau_n$ & $35.0~\mathrm{ns}$ & Characteristic electron lifetime \\
$\tau_p$ & $22.5~\mathrm{ns}$ & Characteristic hole lifetime \\
$n_i$ & $5\times10^{11}~\mathrm{cm^{-3}}$ & Intrinsic carrier concentration \\
$p_0$ & $10^{15}~\mathrm{cm^{-3}}$ & Equilibrium hole concentration \\
$n_0$ & $n_i^2/p_0$ & Equilibrium electron concentration \\
$E_g$ & $1.12~\mathrm{eV}$ & Silicon bandgap \\
$E_i$ & $0.56~\mathrm{eV}$ & Intrinsic Fermi level \\
$E_t$ & $0.66~\mathrm{eV}$ & Trap energy level \\
$T_0$ & $293.15~\mathrm{K}$ & Equilibrium temperature \\
$k_B$ & $8.61\times10^{-5}~\mathrm{eV\,K^{-1}}$ & Boltzmann constant \\
$n_1$ &
$n_i\exp[(E_t-E_i)/(k_BT_0)]$ &
Electron density for $E_F=E_t$ \\
$p_1$ &
$n_i\exp[-(E_t-E_i)/(k_BT_0)]$ &
Hole density for $E_F=E_t$ \\
$a_{\mathrm{SRH}}$ &
$\dfrac{\tau_p+\tau_n}{\tau_p(n_1+n_0)+\tau_n(p_1+p_0)}$ &
SRH numerator coefficient \\
$b_{\mathrm{SRH}}$ &
$\dfrac{1}{n_0+p_0}$ &
SRH denominator coefficient \\
$\tau_0$ &
$\tau_p\dfrac{n_0+n_1}{n_0+p_0}
+\tau_n\dfrac{p_0+p_1}{n_0+p_0}$ &
Low-injection lifetime \\
\hline
\end{tabular}
\end{table}

\subsection{Rational thermal-transfer-function model}

The thermal dynamics were first represented by fitting the thermal transfer
function \(H_{\mathrm{th}}(s)\), which maps absorbed power to effective
temperature rise, using a passive pole expansion:
\[
H_{\mathrm{th}}(s)
\simeq
\sum_{k=1}^{N_p}
\frac{c_k}{s+\lambda_k},
\qquad
\lambda_k>0 .
\]
Here \(s\) is the Laplace variable, \(N_p\) is the number of thermal poles,
\(\lambda_k\) is the decay rate of the \(k\)-th thermal pole, and \(c_k\) is the
corresponding residue. The positivity of \(\lambda_k\) ensures that each pole is
stable. Introducing a thermal state \(z_k(t)\) for each pole gives
\[
\frac{dz_k}{dt}
=
-\lambda_k z_k
+
c_kP_{\mathrm{abs}}(t),
\]
with the effective temperature given by
\[
T_{\mathrm{eff}}(t)=\sum_{k=1}^{N_p}z_k(t).
\]
The coupled reduced-order state is therefore
\[
\left[N,z_1,z_2,\ldots,z_{N_p}\right].
\]
The carrier density \(N(t)\) accounts for free-carrier absorption and
free-carrier dispersion, while the thermal states \(z_k(t)\) account for the
distributed thermal memory of the device.

The pole expansion used \(N_p=48\) logarithmically spaced poles between
\(20~\mathrm{Hz}\) and \(10^{10}~\mathrm{Hz}\). The fit was performed up to
\(10^{10}~\mathrm{Hz}\), and the DC value was extrapolated from the first eight
low-frequency points. Positive residues were enforced so that the fitted
thermal response remained passive. Relative weighting was used in the fit to
avoid overemphasizing only the low-frequency magnitude of the thermal response.

\subsection{Impulse-response thermal model}

As an alternative to the pole expansion, the thermal transfer function
\(H_{\mathrm{th}}(\omega)\) was transformed into a causal impulse response
\(h_{\mathrm{th}}(t)\). In this representation, the effective temperature is
computed by convolution with the absorbed-power history:
\[
T_{\mathrm{eff}}(t)
=
\int_0^t
h_{\mathrm{th}}(t-t')P_{\mathrm{abs}}(t')\,dt' .
\]
Here \(t'\) is the integration variable representing previous times. On the
discrete simulation grid, this convolution is written as
\[
T_{\mathrm{eff}}[n]
=
\sum_{m=0}^{M}K[m]P_{\mathrm{abs}}[n-m],
\]
where \(n\) is the current time index, \(m\) is the history index, \(K[m]\) is
the discrete thermal kernel, and \(M\) is the number of retained kernel samples.

A first-order-hold discretization was used to construct the finite kernel. The
reference kernel length was \(500~\mu\mathrm{s}\), and candidate truncation
times from \(30~\mu\mathrm{s}\) to \(250~\mu\mathrm{s}\) were tested. The
critical truncation time was selected by requiring both the signed and absolute
omitted-tail errors to be below \(5\times10^{-3}\). The final simulations used
\(1.75\) times this critical truncation time. To remove same-sample thermal
feedback in the discrete update, the \(K[0]\) contribution was shifted to the
next time sample. This preserves the total kernel sum while imposing a
one-time-step delay on that contribution.

\subsection{Numerical implementation}

The numerical parameters used in the reduced-order simulations are summarized
in Table~\ref{tab:numerical_parameters}. In the rational-pole model, the coupled
carrier--thermal system was integrated using the Radau method. Radau was chosen
because the problem is stiff: the thermal pole expansion spans many decades in
frequency, while the carrier dynamics and nonlinear cavity feedback can evolve
on nanosecond time scales. Explicit integration would therefore require
extremely small time steps for stability. The implicit Radau method provides a
high-order stiff integration scheme, while the maximum step size was restricted
to match the COMSOL comparison grid.

\begin{table}[htbp]
\centering
\caption{Numerical parameters used in the reduced-order simulations.}
\label{tab:numerical_parameters}
\small
\begin{tabular}{lll}
\hline
Parameter & Value & Purpose \\
\hline
$N_p$ & $48$ & Number of thermal poles \\
Pole range & $20~\mathrm{Hz}$--$10^{10}~\mathrm{Hz}$ & Rational-fit bandwidth \\
$F_{\mathrm{fit,max}}$ & $10^{10}~\mathrm{Hz}$ & Maximum fitted frequency \\
$N_{\mathrm{DC}}$ & $8$ & Points used for DC extrapolation \\
Solver & Radau & Stiff implicit ODE solver \\
Relative tolerance & $10^{-7}$ & ODE accuracy control \\
Absolute tolerance & $10^{-9}$ & ODE accuracy control \\
Maximum step & $0.5~\mathrm{ns}$ & Time-step restriction \\
$N_{\mathrm{scale}}$ & $10^{22}~\mathrm{m^{-3}}$ & Carrier state scaling \\
$T_{\mathrm{scale}}$ & $25~\mathrm{K}$ & Thermal state scaling \\
FOH reference length & $500~\mu\mathrm{s}$ & Impulse-tail reference \\
FOH candidate range & $30$--$250~\mu\mathrm{s}$ & Truncation search range \\
Tail-error target & $5\times10^{-3}$ & Kernel truncation criterion \\
FOH multiplier & $1.75$ & Final truncation factor \\
TR-BDF2 parameter & $2-\sqrt{2}$ & Implicit FOH update \\
\hline
\end{tabular}
\end{table}

The impulse-response simulation used a TR-BDF2 update for the coupled
carrier--thermal-history step. TR-BDF2 combines a trapezoidal-rule stage with a
backward-differentiation stage, giving good stability for stiff nonlinear
systems while retaining second-order accuracy. In this work, the stage
parameter was chosen as
\[
\gamma_{\mathrm{TRBDF2}}=2-\sqrt{2}.
\]
The unknowns \(N\) and \(T_{\mathrm{eff}}\) were solved implicitly at each stage
because the absorbed power depends on the instantaneous cavity detuning,
carrier density, and temperature through the nonlinear relations for
\(\gamma\), \(\Delta\), \(U\), and \(P_{\mathrm{abs}}\).

\subsection{Assignment of models to the main-text results}

The main-text figures do not all represent the same optical operating point.
Their model assignments are summarized in Table~\ref{tab:main_figure_mapping}.
In particular, the different self-pulsation periods in Figs.~2 and 4 of the
main text arise from differences in input power, cold-cavity detuning, and
carrier dynamics; they are not two simulations of an identical operating
condition.

\begin{table}[htbp]
\centering
\caption{Model and operating-condition assignment for the main-text results.}
\label{tab:main_figure_mapping}
\small
\begin{tabular}{p{0.15\textwidth}p{0.34\textwidth}p{0.39\textwidth}}
\hline
Main-text result & Dynamical model & Thermal model and operating condition \\
\hline
Figure~1 & Feed-forward TE$_1$ heating; no cavity feedback & Centroid-sampled impulse response compared with the globally weighted transient temperature \\
Figure~2 & Detailed cavity dynamics with SRH lifetime and separate FCD contributions & Rational pole fit and FOH impulse model; power and detuning taken from the detailed COMSOL case \\
Table~2 & Simplified Borghi-based dynamics with constant lifetime and one effective FCD term & Rational pole fit; input power is varied according to the cases listed in Table~\ref{tab:supp_table2_input_cases} \\
Figure~3 and Sec.~6 & Same simplified dynamics and rational pole fit as the Table~2 comparisons & Centroid and globally weighted TTFs compared at the same simplified cavity operating condition; the displayed transient uses the \(12~\mathrm{dBm}\) case \\
Figure~4 & Simplified dynamics with constant lifetime and one effective FCD term & FOH impulse kernels of different durations; \(12~\mathrm{dBm}\) exported input trace, \(\Delta\nu_0=0\), and a \(900~\mu\mathrm{s}\) simulation window \\
\hline
\end{tabular}
\end{table}

\section{Simulation cases used for Table 2}
\label{sec:supp_table2_cases}

Table~2 of the main paper summarizes the agreement between the COMSOL transient
simulations and the reduced-order rational pole-fit model for different optical
input conditions. These cases use the simplified cavity dynamics employed by
Borghi \textit{et al.}~\cite{borghi2021modeling}, rather than the detailed model
used for Fig.~2 of the main text. The optical-field equations, loss terms, and
absorbed-power definition remain unchanged, but the carrier lifetime is taken
to be constant and free-carrier dispersion is represented by a single effective
term:
\[
\frac{dN}{dt}=g_{\mathrm{TPA}}U^2-\frac{N}{\tau_{\mathrm{fc}}},
\]
\[
\Delta
=
2\pi\Delta\nu_0
+
\omega_0
\left[
-\frac{1}{n_{\mathrm{Si}}}\frac{dn}{dT}T_{\mathrm{eff}}
-\frac{\sigma_{\mathrm{eff}}N}{n_{\mathrm{Si}}}
\right].
\]
The separate carrier contributions and density-dependent SRH lifetime introduced
in Sec.~\ref{sec:supp_reduced_model} are therefore not used in these simplified
cases. The optical input amplitude and cold-cavity detuning are set independently
for each case, as specified by the corresponding COMSOL model and by the input
definitions below.

All Table~2 simulations used the same thermal transfer function,
and the same stable rational pole-fit representation of \(H_{\mathrm{th}}\). The fitted
thermal model used \(N_p=48\) poles over the range
\(20~\mathrm{Hz}\)--\(10^{10}~\mathrm{Hz}\), with positive residues, relative
weighting, and DC extrapolation from the first eight low-frequency points. The
coupled carrier--thermal system was integrated using the Radau solver with the
same state scaling used in Sec.~\ref{sec:supp_reduced_model}.

The sensitivity comparison in Sec.~6 and Fig.~3 of the main text uses this same
simplified cavity model and the same rational pole-fit procedure. Within that
comparison, \(p_{\mathrm{in}}(t)\), \(\Delta\nu_0\), and all carrier parameters
are held fixed; only the thermal transfer function is changed from the
centroid-sampled response \(H(\mathbf r_c,\omega)\) to the globally weighted
response \(H_{\mathrm{eff}}(\omega)\). The displayed transient corresponds to
the \(12~\mathrm{dBm}\) case. The input-power dependence in Table~2 is produced
by varying \(p_{\mathrm{in}}(t)\) as specified below.

The different entries in Table~2 differ in the optical input waveform and, where
applicable, the simulation window. Two
types of input were used. In the COMSOL-driven cases, the input field amplitude
\(p_{\mathrm{in}}(t)\) was read directly from the COMSOL transient export and
interpolated onto the ODE solver grid. The corresponding optical input power was
defined as
\[
P_{\mathrm{in}}(t)=p_{\mathrm{in}}^2(t).
\]
For these cases, the reported mean power and modulation depth were computed
from the exported input trace:
\[
\overline{P}_{\mathrm{in}}
=
\left\langle P_{\mathrm{in}}(t)\right\rangle,
\]
\[
P_{\mathrm{amp}}
=
\frac{P_{\max}-P_{\min}}{2},
\]
\[
\overline{P}_{\mathrm{dBm}}
=
10\log_{10}
\left(
\frac{\overline{P}_{\mathrm{in}}}{1~\mathrm{mW}}
\right),
\]
and
\[
m_{\%}
=
100\frac{P_{\mathrm{amp}}}{\overline{P}_{\mathrm{in}}}.
\]
Thus the modulation is reported as
\[
\overline{P}_{\mathrm{dBm}} \ \pm \ m_{\%}.
\]

In the analytically modulated cases, the COMSOL report function
\(\mathrm{P_{in}}(t)\) defined the optical input power in watts. The reduced
model then used
\[
p_{\mathrm{in}}(t)=\sqrt{\mathrm{P_{in}}(t)}.
\]
The imposed power modulation was
\[
P_{\mathrm{in}}(t)
=
\overline{P}_{\mathrm{in}}
\left[
1+
m\sin
\left(
2\pi\frac{t}{T_{\mathrm{mod}}}
\right)
\right],
\]
where \(T_{\mathrm{mod}}\) is the modulation period and \(m\) is the fractional
modulation amplitude. The reference 5 dBm power was
\[
P_{5\mathrm{dBm}}
=
10^{-3}10^{5/10}~\mathrm{W}
=
3.162~\mathrm{mW}.
\]

The input cases are summarized in Table~\ref{tab:supp_table2_input_cases}.

\begin{table}[htbp]
\centering
\caption{Input definitions for the validation cases reported in Table~2 of the main paper. LC denotes a limit-cycle case, C a constant-input case without self-pulsation, and M a modulated-input case.}
\label{tab:supp_table2_input_cases}
\small
\begin{tabular}{p{0.34\textwidth}p{0.20\textwidth}p{0.16\textwidth}p{0.16\textwidth}}
\hline
Main-text case & Input definition & Mean power & Modulation amplitude \\
\hline
\(12~\mathrm{dBm}\) (LC) & COMSOL \(p_{\mathrm{in}}(t)\) & From exported trace & From exported trace \\
\(11~\mathrm{dBm}\) (LC) & COMSOL \(p_{\mathrm{in}}(t)\) & From exported trace & From exported trace \\
\(10~\mathrm{dBm}\) (LC) & COMSOL \(p_{\mathrm{in}}(t)\) & From exported trace & From exported trace \\
\(5~\mathrm{dBm}\) (C) & COMSOL \(p_{\mathrm{in}}(t)\) & From exported trace & From exported trace \\
\(4~\mathrm{dBm}\) (C) & COMSOL \(p_{\mathrm{in}}(t)\) & From exported trace & From exported trace \\
\(12~\mathrm{dBm}\) (LC), \(274~\mu\mathrm{s}\) window & COMSOL \(p_{\mathrm{in}}(t)\) & From exported trace & From exported trace \\
\(2~\mathrm{dBm}\,\pm100\%\) at \(1~\mathrm{GHz}\) (M) & Analytical \(P_{\mathrm{in}}(t)\) & \(1.99~\mathrm{dBm}\) & \(\pm100\%\) \\
\(4~\mathrm{dBm}\,\pm12.5\%\) at \(100~\mathrm{MHz}\) (M) & Analytical \(P_{\mathrm{in}}(t)\) & \(4.03~\mathrm{dBm}\) & \(\pm12.5\%\) \\
\(4~\mathrm{dBm}\,\pm12.5\%\) at \(100~\mathrm{kHz}\) (M) & Analytical \(P_{\mathrm{in}}(t)\) & \(4.03~\mathrm{dBm}\) & \(\pm12.5\%\) \\
\(4~\mathrm{dBm}\,\pm12.5\%\) at \(100~\mathrm{Hz}\) (M) & Analytical \(P_{\mathrm{in}}(t)\) & \(4.03~\mathrm{dBm}\) & \(\pm12.5\%\) \\
\hline
\end{tabular}
\end{table}

For the 1 ns modulation case, the imposed input power was
\[
P_{\mathrm{in}}(t)
=
0.5P_{5\mathrm{dBm}}
\left[
1+
\sin
\left(
2\pi\frac{t}{1~\mathrm{ns}}
\right)
\right].
\]

\begin{figure}[H]
\centering

\begin{subfigure}[t]{0.48\textwidth}
    \centering
    \includegraphics[width=\linewidth]{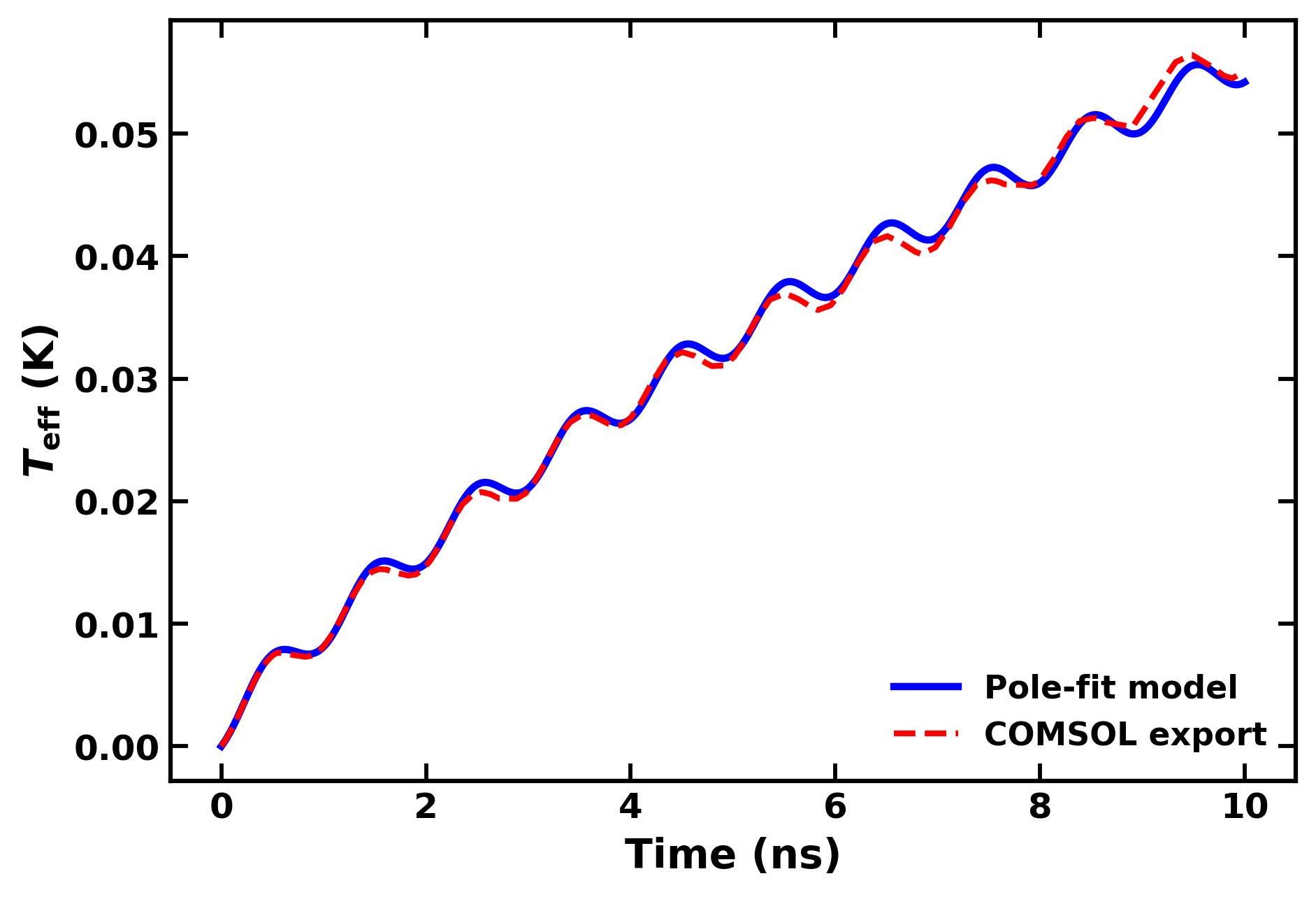}
    \caption{\(2~\mathrm{dBm}\,\pm100\%\) at \(1~\mathrm{GHz}\) (M); the unrounded mean power is \(1.99~\mathrm{dBm}\).}
    \label{fig:table2_1ns}
\end{subfigure}
\hfill
\begin{subfigure}[t]{0.48\textwidth}
    \centering
    \includegraphics[width=\linewidth]{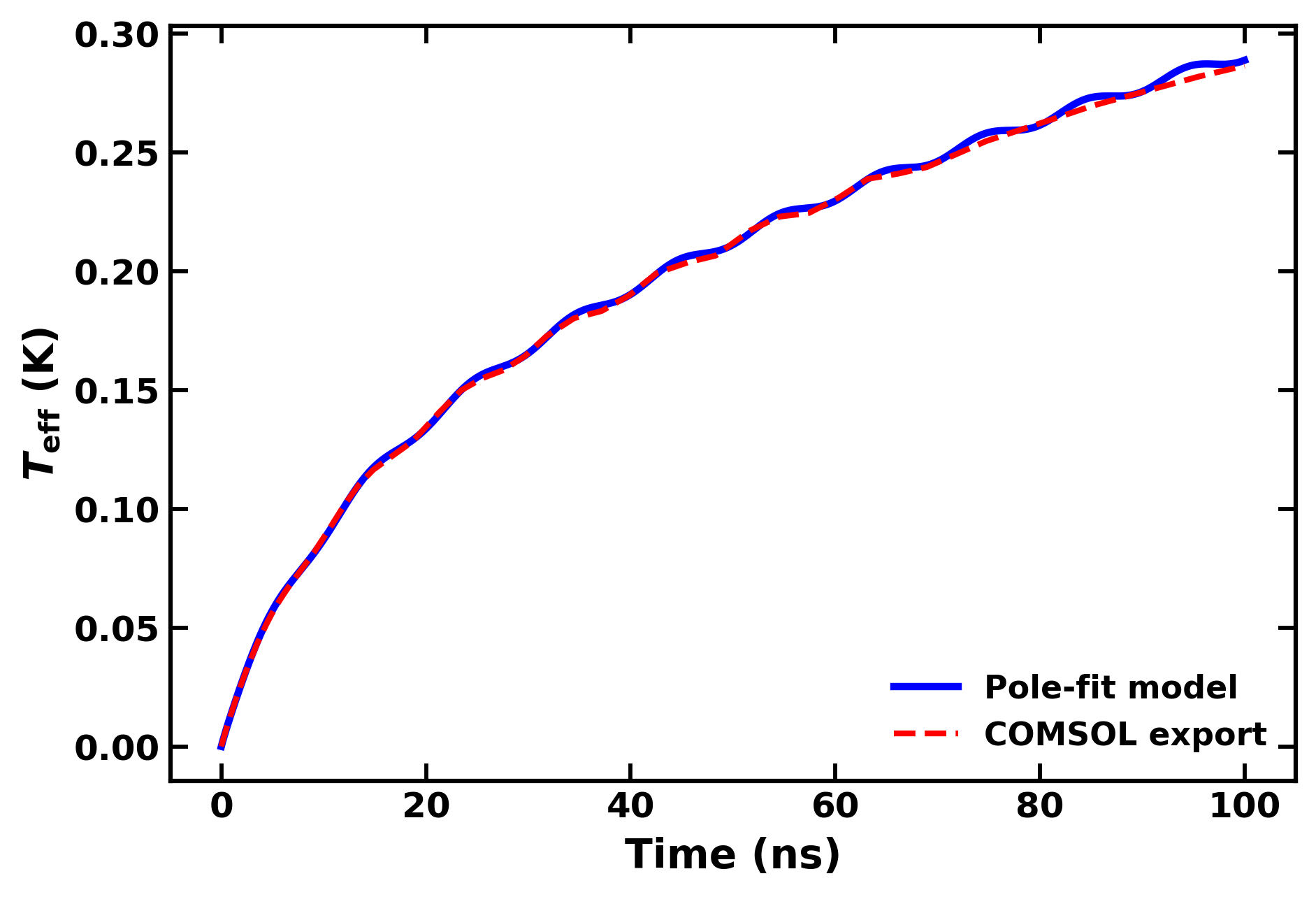}
    \caption{\(4~\mathrm{dBm}\,\pm12.5\%\) at \(100~\mathrm{MHz}\) (M); the unrounded mean power is \(4.03~\mathrm{dBm}\).}
    \label{fig:table2_10ns}
\end{subfigure}

\vspace{0.8em}

\begin{subfigure}[t]{0.48\textwidth}
    \centering
    \includegraphics[width=\linewidth]{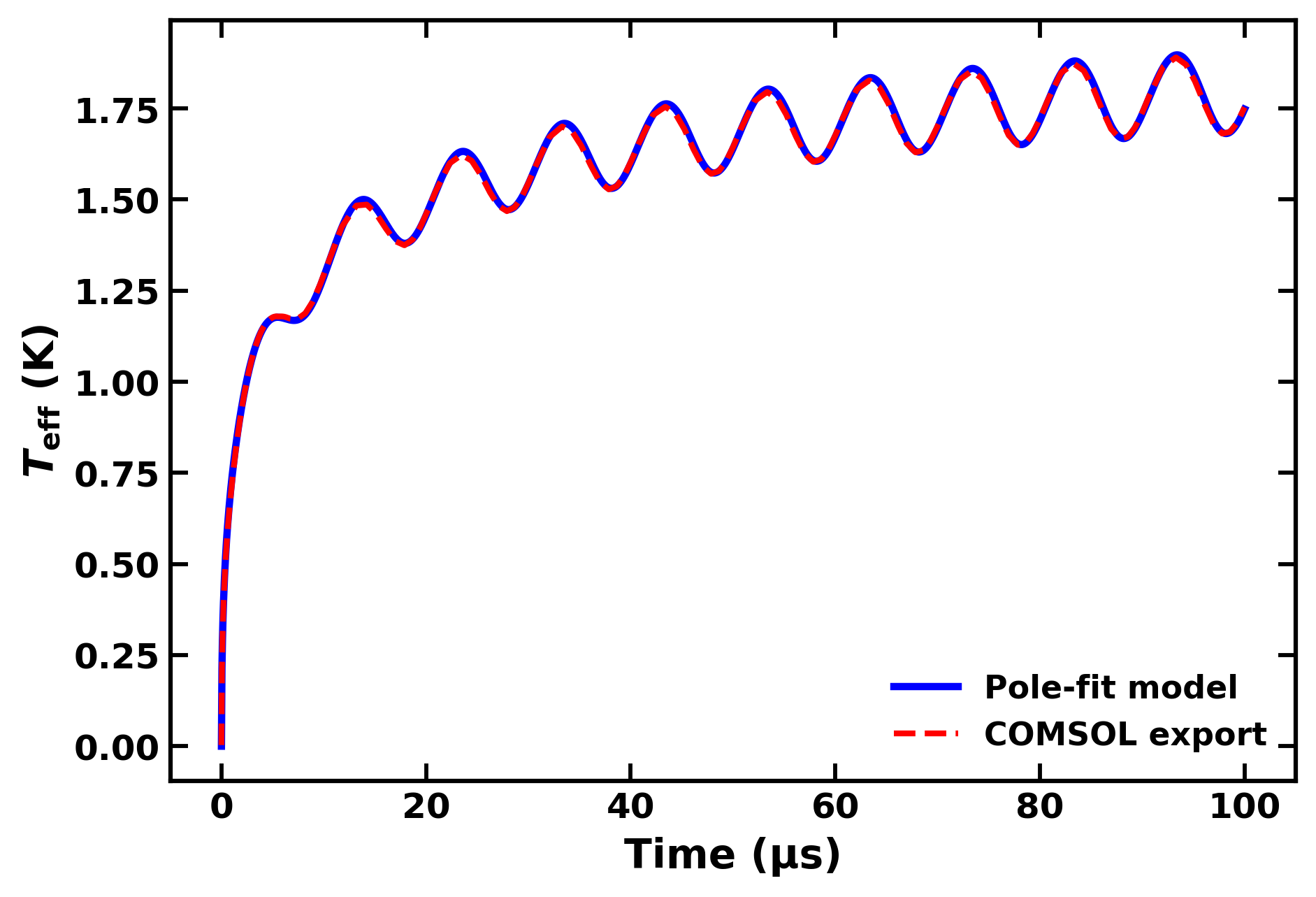}
    \caption{\(4~\mathrm{dBm}\,\pm12.5\%\) at \(100~\mathrm{kHz}\) (M); the unrounded mean power is \(4.03~\mathrm{dBm}\).}
    \label{fig:table2_10us}
\end{subfigure}
\hfill
\begin{subfigure}[t]{0.48\textwidth}
    \centering
    \includegraphics[width=\linewidth]{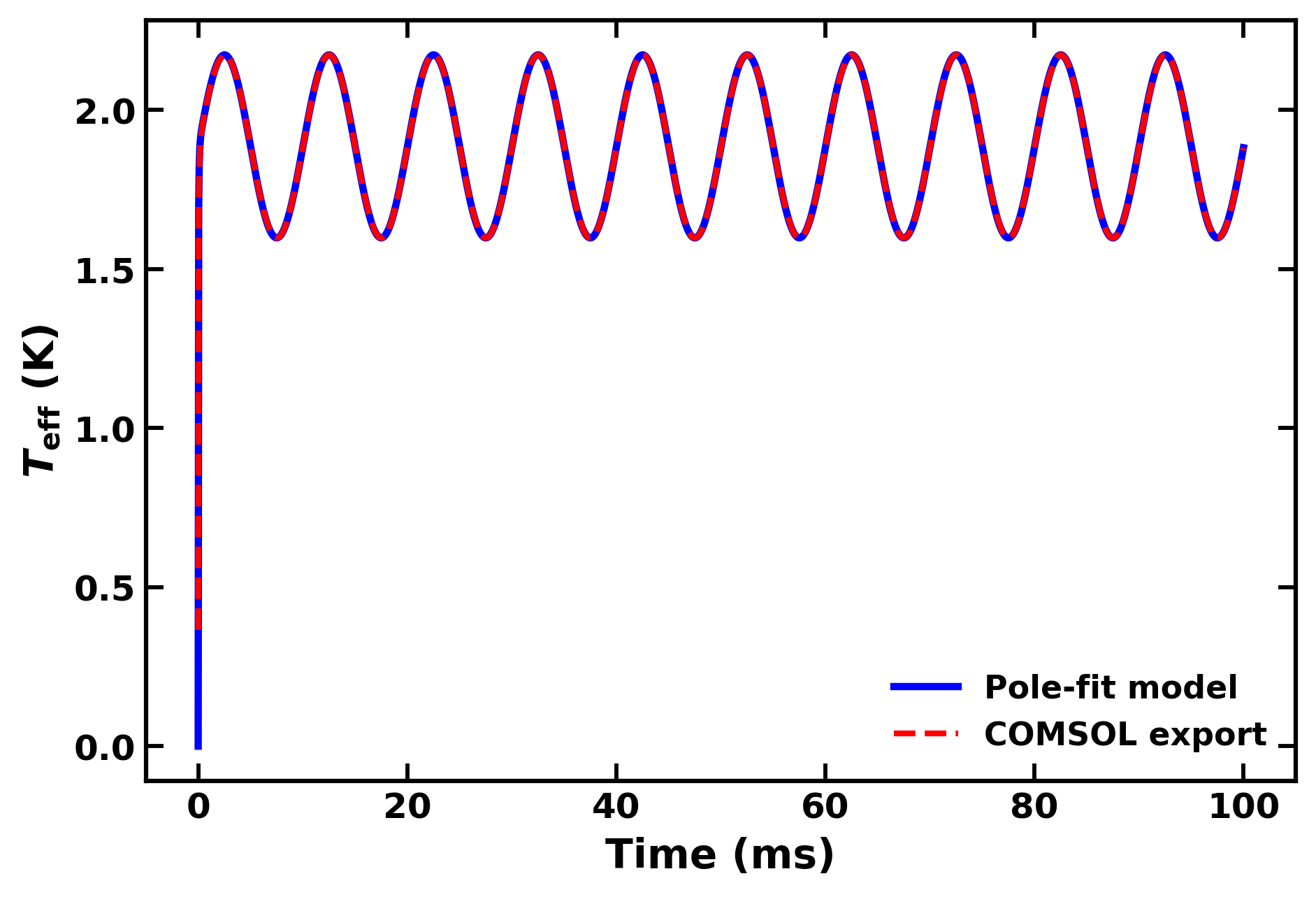}
    \caption{\(4~\mathrm{dBm}\,\pm12.5\%\) at \(100~\mathrm{Hz}\) (M); the unrounded mean power is \(4.03~\mathrm{dBm}\).}
    \label{fig:table2_10ms}
\end{subfigure}

\vspace{0.8em}

\begin{subfigure}[t]{0.48\textwidth}
    \centering
    \includegraphics[width=\linewidth]{figures/Journal_Powers/12_dBm_cavity.png}
    \caption{Constant-input \(12~\mathrm{dBm}\) limit-cycle case (LC).}
    \label{fig:table2_12dbm}
\end{subfigure}
\hfill
\begin{subfigure}[t]{0.48\textwidth}
    \centering
    \includegraphics[width=\linewidth]{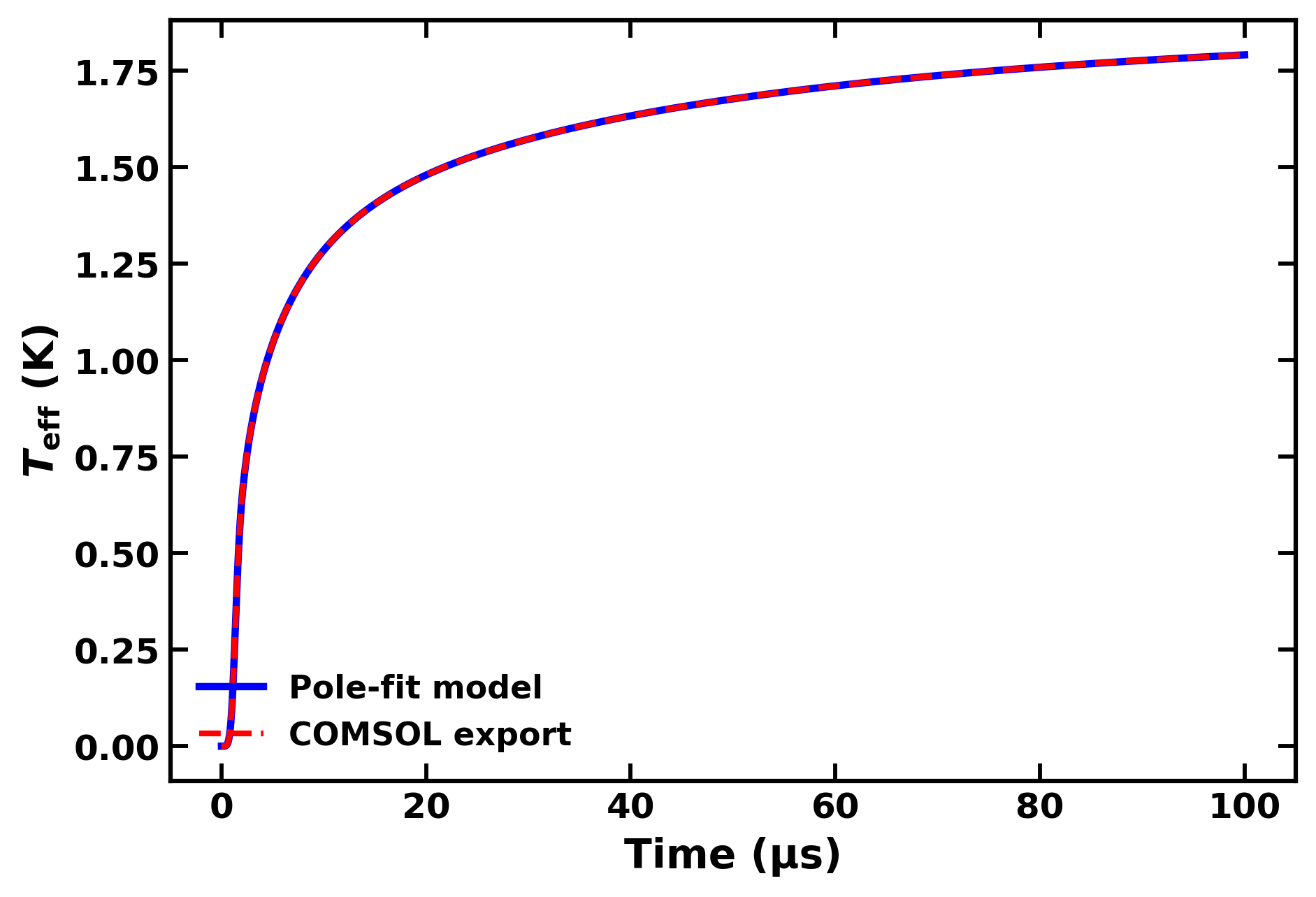}
    \caption{Constant-input \(4~\mathrm{dBm}\) case (C).}
    \label{fig:table2_4dbm}
\end{subfigure}

\caption{
Representative comparisons between COMSOL transient simulations and the
reduced-order rational pole-fit model for the cases summarized in Table~2 of
the main paper. Panels (a)--(d) show the
modulated-input cases, specified by their mean optical input power and relative
power modulation amplitude. Panels (e) and (f) show constant-input simulations
at \(12~\mathrm{dBm}\) and \(4~\mathrm{dBm}\), respectively.
In each panel, the reduced-order model is compared directly against the COMSOL
reference for the corresponding transient response.
}
\label{fig:table2_all_cases}
\end{figure}

Therefore, the mean power was \(0.5P_{5\mathrm{dBm}}\), corresponding to
\(1.99~\mathrm{dBm}\), and the modulation amplitude was \(\pm100\%\).

For the 10 ns, 10 \(\mu\mathrm{s}\), and 10 ms modulation cases, the imposed
input power was
\[
P_{\mathrm{in}}(t)
=
0.8P_{5\mathrm{dBm}}
\left[
1+
0.125
\sin
\left(
2\pi\frac{t}{T_{\mathrm{mod}}}
\right)
\right].
\]
Therefore, the mean power was \(0.8P_{5\mathrm{dBm}}\), corresponding to
\(4.03~\mathrm{dBm}\), and the modulation amplitude was \(\pm12.5\%\).

For each case, the reduced-model outputs were evaluated on the same time grid
as the COMSOL transient export. The model generated
\(T_{\mathrm{eff}}\), \(N\), \(a_r\), \(a_i\), and \(U\); Table~2 of the main
text reports the percentage RMSE of \(T_{\mathrm{eff}}\), computed as
\[
\mathrm{RMSE}_{\%}
=
100
\frac{
\sqrt{
\left\langle
\left(
x_{\mathrm{model}}-x_{\mathrm{COMSOL}}
\right)^2
\right\rangle
}
}{
\sqrt{
\left\langle
x_{\mathrm{COMSOL}}^2
\right\rangle
}
}.
\]

Figure~\ref{fig:table2_all_cases} provides representative temperature traces
and their comparison with the corresponding COMSOL simulations.

\section{Impulse-kernel truncation and convergence analysis}
\label{sec:supp_impulse_convergence}

The convergence study corresponding to Fig.~4 of the main text used the
simplified cavity dynamics described in Sec.~\ref{sec:supp_table2_cases}, not
the detailed SRH model used for Fig.~2. It used the exported \(12~\mathrm{dBm}\)
input trace, zero cold-cavity detuning, and a \(900~\mu\mathrm{s}\) simulation
window. These differences in power, detuning, and carrier dynamics account for
the different absolute self-pulsation periods in Figs.~2 and 4 of the main text.

The convergence outputs for the impulse-response model were generated from a
single long-duration reference kernel. Before inversion, the sampled thermal
transfer function was sorted by frequency and its zero-frequency value was
estimated from the first eight low-frequency points. Specifically, the real
part was extrapolated linearly as a function of \(f^2\), while the imaginary
part was constrained to vanish at DC. The transfer function was then
interpolated onto a uniform frequency grid using piecewise-cubic Hermite
interpolation. A cosine taper was applied over the upper portion of the
frequency range to suppress ringing caused by an abrupt spectral cutoff.

Because the inverse discrete Fourier transform represents a periodic time
record, the frequency spacing was chosen so that the corresponding time window
was at least
\[
T_{\mathrm{IFFT}}
\geq
\max\left(
3T_{\mathrm{ref}},
T_{\mathrm{ref}}+300~\mu\mathrm{s}
\right),
\]
where \(T_{\mathrm{ref}}=500~\mu\mathrm{s}\) is the reference-kernel duration.
This guard interval prevents the unresolved impulse-response tail from wrapping
around into the beginning of the reconstructed kernel. The maximum reconstructed
frequency was \(5~\mathrm{GHz}\).

The fine-grid impulse response was converted to the transient-simulation grid
using a first-order-hold representation of the absorbed power. For a coarse
time step \(\Delta t\), the FOH kernel coefficient at delay \(r\Delta t\) was
calculated by projecting the fine impulse response onto the triangular basis
\[
\Lambda_r(\tau)
=
\max\left[
1-\frac{\left|\tau-r\Delta t\right|}{\Delta t},
0
\right].
\]
Thus, if \(h_q\) denotes the discrete fine-grid impulse-response weight at
\(\tau_q\), the coarse coefficient is
\[
K_{\mathrm{ref}}[r]
=
\sum_q h_q\Lambda_r(\tau_q).
\]
The zero-delay coefficient was transferred to the first delayed coefficient,
\[
K_{\mathrm{ref}}[1]\leftarrow
K_{\mathrm{ref}}[1]+K_{\mathrm{ref}}[0],
\qquad
K_{\mathrm{ref}}[0]\leftarrow0,
\]
thereby removing same-sample thermal feedback and imposing a one-time-step
delay on this coefficient. This operation preserves the total kernel sum and
hence its DC thermal gain.

To quantify the error introduced by finite kernel duration, the
\(500~\mu\mathrm{s}\) reference kernel was divided into retained and omitted
portions at each candidate cutoff \(T_c\). If \(M_c\) is the number of samples
retained at that cutoff, the signed and absolute omitted-tail measures were
defined in Eqs.~(15) and (16) of the main text.

The signed metric measures the remaining error in the integrated, or DC,
thermal response, whereas the absolute metric prevents cancellation between
positive and negative portions of the omitted tail. Candidate cutoffs between
\(30~\mu\mathrm{s}\) and \(250~\mu\mathrm{s}\), in increments of
\(10~\mu\mathrm{s}\), were evaluated. The critical truncation \(T_{\mathrm{crit}}\)
was defined as the first candidate satisfying
\[
\epsilon_{\mathrm{signed}}\leq5\times10^{-3},
\qquad
\epsilon_{\mathrm{abs}}\leq5\times10^{-3}.
\]
This criterion gave \(T_{\mathrm{crit}}=180~\mu\mathrm{s}\). The longest
kernel used in the nonlinear comparisons was therefore
\(1.75T_{\mathrm{crit}}=315~\mu\mathrm{s}\).

To verify that the kernel-tail criterion also produced converged nonlinear
dynamics, simulations were performed using truncation times
\[
T_c=
\left\{
0.6,\;0.8,\;1.0,\;1.25,\;1.5,\;1.75
\right\}T_{\mathrm{crit}}.
\]
All cases used the same time grid, physical parameters, initial conditions, and
input waveform. The simulation interval was \(900~\mu\mathrm{s}\), and the time
step was taken from the COMSOL transient export. The exported input-field
amplitude was linearly interpolated onto this grid; if the reduced-model
simulation extended beyond the available COMSOL trace, its final input value
was held constant.

For each accepted time step, the newly calculated absorbed power was stored in
the thermal memory by multiplying it by the retained kernel:
\[
T_{\mathrm{eff}}[n]
=
\sum_{m=0}^{M}K[m]P_{\mathrm{abs}}[n-m].
\]
Thus, every absorbed-power sample contributes to the temperature at later
times, with \(K[m]\) specifying how much of that contribution remains after
\(m\) time steps.

TR-BDF2 also evaluates the system at an intermediate time,
\[
t_{n+\gamma}=t_n+\gamma\Delta t,
\qquad \gamma=2-\sqrt{2}.
\]
Because this intermediate time lies between two kernel samples, the required
kernel value was obtained by linear interpolation between the neighboring
values of \(K[m]\). This provides the thermal history at the intermediate
TR-BDF2 stage without introducing a separate, finer time grid. Since the
same-sample coefficient was set to \(K[0]=0\), the temperature at each
stage depends only on previously accepted absorbed-power values and not on an
instantaneous same-sample thermal response. Carrier density and effective temperature
were then solved simultaneously at both implicit stages, accounting for the
instantaneous dependence of absorbed power on the nonlinear cavity state.

The resulting outputs were the effective-temperature rise
\(T_{\mathrm{eff}}(t)\), carrier density \(N(t)\), and absorbed power
\(P_{\mathrm{abs}}(t)\) for each truncation. Where available, the COMSOL
transient was included in the temperature and carrier-density plots as an
independent reference over its exported time interval. The omitted-tail plot
was generated from the reverse cumulative signed and absolute sums of the
reference kernel, with the threshold levels and tested truncation times marked
explicitly.

Self-pulsation periods were obtained from consecutive maxima of
\(T_{\mathrm{eff}}(t)\). A peak was retained if its prominence exceeded
\(25\%\) of the full temperature excursion, and adjacent peaks were required
to be separated by at least \(5~\mu\mathrm{s}\). If the detected peak times are
\(t^{\mathrm{pk}}_j\), the period assigned to oscillation \(j\) is
\[
\tau_j=t^{\mathrm{pk}}_{j+1}-t^{\mathrm{pk}}_j.
\]
Plotting \(\tau_j\) against oscillation number reveals both the transient
evolution of the self-pulsation period and its sensitivity to kernel
truncation.

Finally, the thermal-history implementation was checked by independently
recomputing the discrete convolution,
\[
T_{\mathrm{check}}[n]
=
\sum_m K[m]P_{\mathrm{abs}}[n-m],
\]
and comparing it with the temperature obtained during the incremental
TR-BDF2 calculation. Maximum and RMS values of
\(T_{\mathrm{eff}}-T_{\mathrm{check}}\), together with the implicit carrier and
thermal equation residuals, were used to verify numerical consistency of each
simulation.

\bibliography{sample}